\documentclass [article]{aa}
\usepackage{graphicx}

 \usepackage[T1]{fontenc}
 \usepackage{textcomp}
 \usepackage{mathptmx}
 \usepackage{courier}
 \usepackage[scaled=0.94]{helvet}
 \usepackage{hyperref}
 \hypersetup{
     final=true,
     pageanchor=true,
     hyperfigures=false,
     colorlinks=true,
     breaklinks=true,
     linkcolor=blue,
     citecolor=blue,
     urlcolor=blue,
     pdfpagemode=UseNone,
     pdfpagelayout=OneColumn,
     pdfview=FitBH,
     pdftitle={Analysis of global and spatial variations of line parameters},
     pdfauthor={Puschmann et al.},
     pdfsubject={Solar Physics}}

\begin{document}

\title{Time series of high resolution photospheric spectra in a quiet region of the sun}
\subtitle{I. Analysis of global and spatial variations of line parameters}

\author{K. Puschmann \inst{1,2,3} \and M. V\'azquez \inst{2} \and
J.A. Bonet \inst{2} \and B. Ruiz Cobo \inst{2} \and A. Hanslmeier\inst{3}}

\offprints{K. Puschmann}

\institute{Universit\"ats-Sternwarte, Geismarlandstr. 11, D-37083 G\"ottingen, Germany\\
\email{kgp@uni-sw.gwdg.de}
\and
Instituto de Astrof\'\i sica de Canarias, C/V\'\i a 
L\'actea s/n,E-38200 La Laguna, Spain\\
\email{mva@ll.iac.es, jab@ll.iac.es, brc@ll.iac.es}
\and 
Institut f\"ur Geophysik, Astronomie und Metereologie, Universit\"atsplatz 5, A-8010 Graz, Austria\\
\email{arnold.hanslmeier@kfunigraz.ac.at}}

\date{Received 26 November 2002; accepted 19 June 2003}
%\maketitle
%\markboth {}{}

\abstract{A 50 min time series of one-dimensional slit-spectrograms, taken in quiet sun at disk centre, observed at the German Vacuum Tower Telescope (Observatorio del Teide), was used to study the global and spatial variations of different line parameters. In order to determine the vertical structure of the photosphere two lines with well separated formation heights have been considered. The data have been filtered of p-modes to isolate the pure convective phenomenon. From our studies of global correlation coefficients and coherence and phase shift analyses between the several line parameters, the following results can be reported. The convective velocity pattern preserves structures larger than $1\,\farcs0$ up to the highest layers of the photosphere ($\sim$ 435 km). However, at these layers, in the intensity pattern only structures larger than $2\,\farcs0$ are still connected with those at the continuum level although showing inverted brightness contrast. This confirms an inversion of temperature that we have found at a height of $\sim$ 140 km. A possible evidence of gravity waves superimposed to the convective motions is derived from the phase shift analysis. We interprete the behaviour of the full width at half maximum and the equivalent width as a function of the distance to the granular borders, as a consequence of enhanced turbulence and/or strong velocity gradients in the intergranular lanes.
}
\titlerunning{Time series of high resolution photospheric spectra}
\maketitle
%
%_________________________________

\section{Introduction}

An important part of our knowledge about solar granulation has been obtained
from the analysis of individual images and/or spectrograms. Actually, the most
detailed images are obtained by applying different on-line and post-facto
techniques (see e.g. Bonet \cite{bonet99} for a summary).
Additionally, one-dimensional (1D) and two-dimensional (2D) high spatial resolution spectrograms ($\approx$ 0\,\farcs5) are used to investigate the dynamical behaviour of granules and intergranular lanes and their penetration through the different layers of the photosphere. A few parameters like continuum intensity, line core intensity, equivalent width and full width at half maximum are often used, in addition to line bisectors, in order to characterise the shape of photospheric lines and to infer important physical parameters. Vertical granule velocities are derived from Doppler shift measurements in the line core.\\

First studies of the evolution of granulation were mainly connected with the
determination of lifetimes. However, a detailed description of the evolution
of individual granules demands data with much better quality than that
required for mere identification. Mehltretter (\cite{mehltretter}) using high 
resolution time series of white light images found two types of behaviour: On time scales shorter than six minutes granules or groups of granules seem to expand. At larger time scales the individual granules fragment or drift. Dialetis et al. (\cite{dialetis}) described the different types of formation (evolution of a small granule, fragmentation of a larger granule, and merging of smaller granules) and disappearance (fading into the background, merging a neighbouring granule or fragmentation) of these structures. Hirzberger et al. (\cite{hirzberger99}) found that fragmentation and merging are the most frequent mechanisms for birth and disappearance, respectively.\\

Concerning the spectroscopic methods, the vertical structure of the convective motions has been studied in detail using correlation and spectral analyses techniques applied to the fluctuations of different line parameters. For such studies different sets of observational data have been used.

\noindent
a) Most investigations have been based on single 1D spectrograms, Nesis et al. (\cite{nesis88}, \cite{nesis92}, \cite{nesis93}), Komm et al. (\cite{komm90},
\cite{komm91a}, \cite{komm91b}, \cite{komm91c}), Hanslmeier et al. (\cite{hanslmeier91a}, \cite{hanslmeier91b}, \cite{hanslmeier93}, \cite{hanslmeier94}), but without filtering properly the p-modes from their data, which strongly masks the convective signature.\\ 
b) Tracking of spectra: Altrock et al. (\cite{altrock}) and Nesis et al. (\cite{nesis99}, \cite{nesis01}) followed the evolution of selected granules during short periods of time.
Keil (\cite{keil}) and Johannesson et al. (\cite{johannesson}) obtained 2D spectral images derived from 1D spectrograms by scanning a certain area of the solar surface. More recently Collados et al. (\cite{collados}) obtained improved data using the Correlation Tracker developed at the Instituto de Astrof\'\i sica de Canarias.\\
c) 2D spectrograms have been obtained by using a Multichannel Subtractive Double Pass (MSDP) spectrograph (Roudier et al. \cite{roudier}; Espagnet et al. \cite{espagnet93}, \cite{espagnet95}) or Fabry-Perot interferometers (Salucci et al. \cite{salucci}; Bendlin \& Volkmer \cite{bendlin};  Krieg et al. \cite{krieg}; Hirzberger et al. \cite{hirzberger01}).\\

From all these investigations the granulation phenomenon can be currently
described as follows: convective overshoot of the plasma flow from the solar
convection zone into the photosphere forms a pattern of bright cellular
elements showing upward motions -- the granules, surrounded by a network of
dark intergranular lanes, where downflow motions are observed. According to
correlation analyses (Deubner \cite{deubner88}; Salucci et
al. \cite{salucci}; Espagnet et al. \cite{espagnet95}) the horizontal
temperature fluctuations associated with these motions decrease rapidly with
increasing height until they vanish. Only the largest granules ($>$ 1\,\farcs5) contribute to the brightness pattern observed above where the brightness contrast is inverted.  Note that the values about the height where the intensity
fluctuations vanish varies a lot in the literature. One finds values from
60 km reported by Kneer et al. (\cite{kneer}), 60-90 km (Espagnet et al.
\cite{espagnet95}) 170 km (Hanslmeier et al. \cite{hanslmeier93}; Komm et
al. \cite{komm91a}) up to 270 km where Bendlin \& Volkmer (\cite{bendlin})
detected brightness signatures of the granulation. The values may differ from
each other due to differences in the method to establish the geometrical
height scale (transformation from $\tau$ to $z$) and in different methods to filter the p-modes. The vertical velocity field persists in the upper layers, but waves (acoustic and gravity) and turbulent motions may also contribute to the velocity fluctuations (Deubner \cite{deubner88}; Deubner \& Fleck \cite{deubner89}; Salucci et al. \cite{salucci}; Espagnet et al. \cite{espagnet95}). An overview on basic properties of solar granulation can be found for instance in Bray et al. (\cite{bray}), Spruit et al. (\cite{spruit}), Muller (\cite{muller99}) and Brandt (\cite{brandt}).\\

In order to get a detailed insight into the spatial and temporal evolution of
granulation, a series of high spatial resolution spectroscopic observations
is required, which is, however, extremely difficult to obtain. Some conditions
should be fulfilled {\em a priori} for this study:\\
a) A time series of spectrograms, long enough to allow an adequate filtering of the p-modes, because the presence of oscillations tends to decorrelate intensity and velocity fluctuations.\\
b) In principle, a 2D coverage of the solar surface is desirable
to sample the full evolution of granular areas. Nevertheless, time series of 1D slit spectra still supply worthwhile information, provided we have an adequate tracking of the solar structures.\\
c) A good seeing stability over the whole time series to allow the dynamical study of individual granules.\\

We analyse a time series of 1D slit spectrograms, recorded at the German
Vacuum Tower Telescope (Observatorio del Teide, Tenerife) which fulfills the above mentioned conditions. In this paper we present a study of the global and spatial variations of different line parameters in a quiet region
at the solar disk centre. In Sect. 2 of this paper we describe the
observations and the process of data reduction. In Sect. 3 the tools for data analyses are discussed, i.e. determination of line parameters and formation height of spectral lines, sub-sonic filtering, coherence and phase spectral analysis, and image segmentation. In Sect. 4 global correlation between several line parameters derived from our spectrograms, for filtered and unfiltered data, are investigated. In addition, the mean values of line parameters calculated in granular and intergranular areas are compared. In order to get a deeper knowledge about the penetration of different structures into the photosphere we perform in Sect. 5 a spatial coherence and phase shift analysis between the several line parameters and we study the dependence of these parameters on the spatial distance to the granular borders. In Sect. 6 we examine two particular granular cells, large and small, respectively, and the results are compared with the previous statistical findings for the whole sample. Finally, we summarise in Sect. 7 the conclusions of this work.

\section {Observations and data reduction}
\label{obsred}
\subsection{Observations}
\label{obs}

\begin{table*}[h,t]
\small
\begin{flushleft}
\begin{tabular}{lllrrrrl}
\hline Line & Element &  $\lambda $ (\AA) & $W_{\lambda}$ (m\AA) &
$P_{\rm exc}$ (eV) & $\log \, gf$ & Reference\\
\hline
(I) & \ion{Fe}{i} & 6494.994 & 165.0 & 2.39 & -- & Moore et al. \cite{moore} \\
(I) & \ion{Fe}{i} & 6494.994 & 162.0 & 2.40 & $-$1.23 & Gadun et al. \cite{gadun97}, \cite{gadun99} \\
(III) & \ion{Fe}{i} & 6496.476 & 69.0 & 4.77 & -- & Moore et al. \cite{moore} \\
(III) & \ion{Fe}{i} & 6496.472 & 67.7 & 4.79 &$-$0.66 & Gadun et al. \cite{gadun97}, \cite{gadun99} \\
\hline
\end{tabular}
\end{flushleft}
\caption[]{Characteristic parameters of the lines analysed. The columns in
order: line, element, wavelength, equivalent width, excitation
potential, $gf$ ($g$ = statistical weight, $f$ = oscillator strength) and reference to literature.}
\label{alexey}
\end{table*}

\begin {figure}[h]
\centering
\includegraphics[width=8cm,height=6cm]{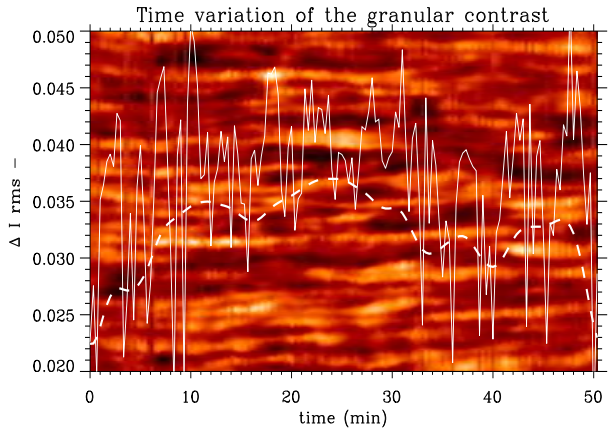}
\caption []{Time variation of the raw $\Delta I_{\rm rms}$ along the slit (solid line) and after filtering of the p-modes (dashed line). Background: image
  of unfiltered continuum intensity in temporal (horizontal axis) and spatial
  direction (vertical axis). The total spatial direction covers a range of
  46\,\farcs6.}
\label{Irms}
\end {figure}

Time series of CCD spectrograms were obtained in July 8, 1993 with the 70 cm
Vacuum Tower Telescope (VTT, see Schr\"oter et al. \cite{schroeter}, for a
description) at the Observatorio del Teide (Tenerife). From the recorded
material, a time series spanning 50 min has been selected for this study, due
to the stability of the seeing conditions. Figure \ref{Irms} shows the time
variation of the continuum $\Delta I_{\rm rms}$ (normalised rms contrast of
the granulation) along the slit. The strong spikes of the unfiltered signal
are related with variations of seeing which also produce very high time-frequency structures (vertical narrow streaks) in the background image of Fig. 1. This information occurs in the $k$-$\omega$ diagram within the domain removed by the p-modes filter. Consequently, the narrow streaks are smoothed out as a by-product of the p-modes filtering. This smoothing also affects the $\Delta I_{\rm rms}$ in the spatial direction (see dashed line in Fig. 1).

Magnetically active regions have been avoided by monitoring the entrance slit
position in  \ion{Ca}{ii} \ion{K}{} slit-jaw images. The observations have been performed at the solar disk centre in order to obtain the vertical velocity
fluctuations. 

We have used a nitrogen cooled CCD camera with 1024 $\times$
1024 pixels. The exposure time was 1.5 s and the interval between two
consecutive spectrograms 20 s. The entrance slit of the spectrograph was set
to 100 $\mu$ (0\,\farcs46) and the CCD images included three photospheric
\ion{Fe}{i} lines  ($\lambda$ 6494.994, 6495.740, 6496.476 \AA ) that had
been studied in previous papers (Hanslmeier et al. 1990) and will be
hereafter referred to as I, II, and III, respectively (see Table \ref{alexey}
and Fig. \ref{filtres}). Line II has been finally excluded from this analysis
due to a blend with a terrestrial water vapour line. A short summary of the
most important properties of the spectral lines studied is given in
Table \ref{alexey}. The pixel size in the focal plane of the spectrograph was
0\,\farcs093 in the spatial direction and 2.06 m\AA\,\,($\approx$ 95.15
m\,s$^{-1}$) in the spectral direction. In spatial direction, the slit covered a total length of $95\,\farcs2$ (1024 pix) although after flatfielding, only the central part between the two reference hairs for positioning (46\,\farcs6 equivalent to 501 pix) was kept for further analyses. From the computed power spectra we estimate the effective spatial resolution achieved as $\sim$ 0\,\farcs5.\\

To follow the time evolution of a physical point on the solar surface one
must ensure a perfect tracking of this point on the spectrograph slit all
the time. Image motion induced by atmospheric turbulence and mechanical
vibrations of the telescope degrade the angular resolution and contaminate
the observation with information stemming from neighbouring points. Similarly,
a drift of the image caused by non-perfect guiding of the telescope leads to a
loss of tracking of the point of interest. To minimise these drawbacks
originating from global image motion, our observations were performed
through a Correlation Tracker developed at the Instituto de Astrof\'\i sica
de Canarias (Bonet et al. \cite{bonet92}; Ballesteros et
al. \cite{ballesteros94}; Bonet et al. \cite{bonet94}; Ballesteros et
al. \cite{ballesteros96}) that was available for the first time in 1993 for
scientific observations. This device, mounted in front of the spectrograph
slit at the VTT, has a tilting mirror which compensated for overall wavefront
tilts -- responsible of the global motion -- at a fast rate. The system allowed a precise stabilisation of a portion of the solar image as well as the tracking of the corresponding area on the solar surface for long periods, following the solar rotation and the motions induced by the large-scale velocity pattern. This Correlation Tracker proved to have a good performance. Small faults in the tracking of the observed area, along the spectrograph slit ($\pm$ 1 px) were corrected {\it a posteriori} by cross-correlating the continuum intensity profiles of adjacent spectrograms in the spatial direction.

\subsection{Flatfielding}
\begin{figure}
\centering
\includegraphics[width=8cm,height=4.5cm]{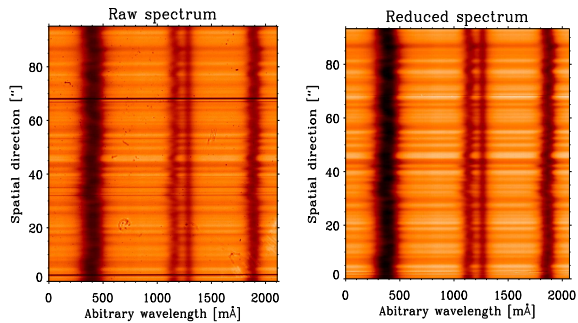}
\caption []{Particular raw spectrum (left panel) and corresponding reduced
  spectrum (right panel). In both panels we can see from left to right: Line I, Line II, the telluric blend, and Line III.}
\label{filtres}
\end{figure}

The flat fields have been obtained in the same spectral region by moving the
pointing of the telescope across the centre of the solar disk. 
The presence of sharp lines in wavelength direction (hereafter scratches) in
the obtained spectrograms (see Fig. \ref{filtres}, left panel, where a particular raw spectrum is presented) and flat fields has required a special procedure to reduce our data, since parts of our further analyses would have been strongly affected by these artifacts.

We have two types of scratches. The first stems from hair lines placed in front of the spectrograph slit (corresponding to the two darkest horizontal lines
visible in the upper and the lower part of the raw spectrum and flat field
source frame, respectively). The second type of scratches results from
dust particles on the spectrograph slit. 

In the following, we present a brief description of the data
reductions we have applied (for a similar procedure see also W\"{o}hl et al. \cite{woehl}):

\begin{itemize}
\item[a)]
Subtraction of the dark current from the particular spectra and flat fields. 
\item[b)]
Construction of the flat field source frame by averaging the individual flats,
taken at the beginning and the end of the observing run.
\item[c)]
The removal of the spectral lines from the flat field source frame requires the division of the flat field source frame by its mean spectral profile averaged in spatial direction. To perform this correction in a proper manner one has to rectify the possible inclination of the spectral lines with respect to the vertical edges of the frame.
\item[d)] 
Splitting up the flat field matrix into two components: {\it slit flat}
(containing the systematic effect induced by dust  and wires on the spectrograph slit) and {\it camera flat} (including the rest of the spurious systematic effects)
\item[e)] 
After dividing each spectrogram by the {\it camera flat}, we have compared by correlation the $y$-positions of the most prominent horizontal scratches in the {\it slit flat} and in each individual spectrogram. From this comparison the optimum shift of the slit flat to match the position of the scratches in the spectrogram has been derived with subpixel precision. Finally, each spectrogram has been divided by the displaced {\it slit flat}. A particular reduced spectrum is presented in the right panel of Fig. \ref{filtres}.
\item[f)] 
Due to some remaining residuals of the hair lines in the reduced spectrum, we
have removed from each spectrogram the upper and lower parts outwards the
hair lines, so that only the central area in the spatial direction is kept for 
further studies.  
\end{itemize}

\subsection{Filtering of noise}
\label{noise}
\begin{figure}
\centering
\includegraphics[width=9cm,height=7cm]{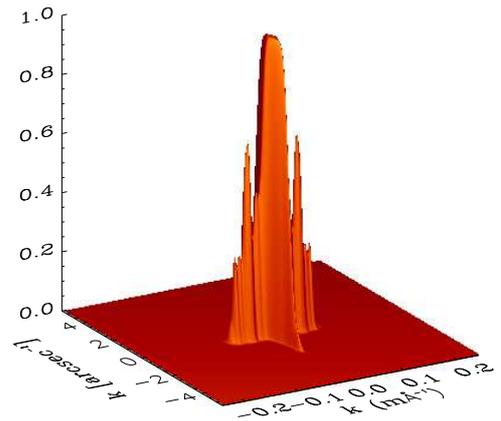}
\caption []{3D representation of the optimum filter to improve the signal to noise ratio and to eliminate the residuals of horizontal scratches in the spectra.}
\label {filter}
\end{figure}

The noise and residuals of some remaining weak horizontal scratches in
our spectra have been filtered out in the Fourier domain by applying a
2D optimum filter that is a generalisation of the 1D filter described by Brault \& White (\cite{brault}) (see also Bonet \cite{bonet99}). 
The mathematical expression for this filter ($\tilde\Phi (s)$) is:
\begin{equation}
\tilde\Phi ({\bf s})=\frac{P_{\rm S}({\bf s})}{P_{\rm S}({\bf s}) + P_{\rm N}({\bf s})},
\label{equ1}
\end{equation}
where {\bf s} is the frequency, $P_{\rm S}({\bf s})$ and $P_{\rm N}({\bf s})$ are smoothed models of the signal and noise power spectra averaged over all
spectrograms. $P_{\rm N}({\bf s})$ has been modelled as a constant value equal to the mean level of the power in the range of high frequencies. The model for
$P_{\rm S}({\bf s})$ has been obtained by smoothing the mean power spectrum using a boxcar of empirically determined dimensions and then subtracting
$P_{\rm N}({\bf s})$. A 3D representation of the resulting filter is shown in Fig. \ref{filter}. 

\begin{figure*}
\centering
\includegraphics[width=14cm,height=7cm]{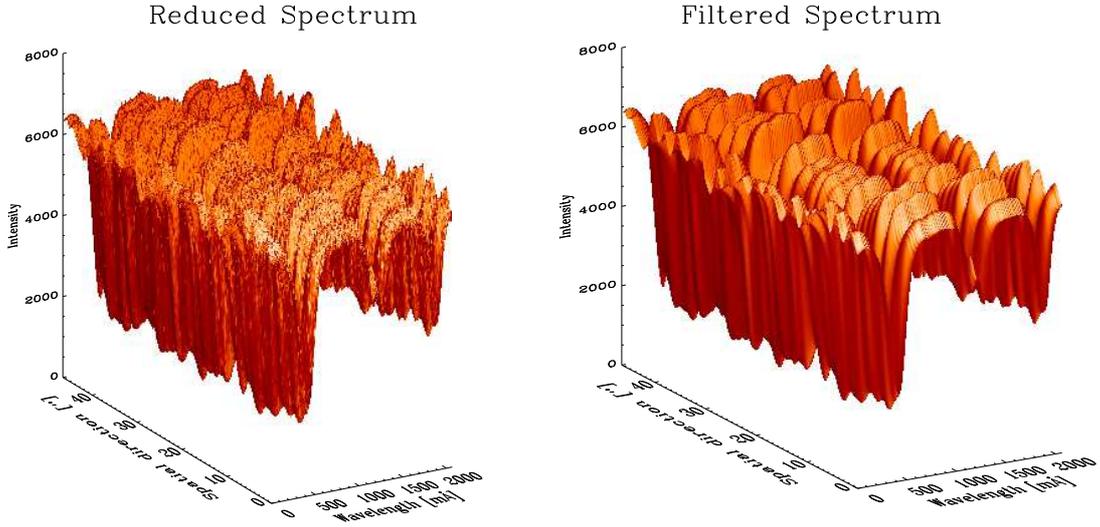}
\caption []{3D representation of an individual spectrogram before (left
panel) and after (right panel) applying the optimum filter to improve the
signal to noise ratio. Spectral-axis: wavelength ($m$\AA) in relative units;
spatial-axis: $arcsec$; vertical-axis: intensity in relative units.}
\label {filterprog1}
\end{figure*}

Figure \ref{filterprog1} shows a particular spectrogram before and after
filtering. 

Each spectrogram has been also corrected for an inclination of the
spectral lines with respect to the vertical edge of the CCD. Finally, the average continuum of each spectrogram has been fitted to the local continuum of the Li\`ege Atlas (Delbouille et al. \cite{delbouille73}) at this wavelength range.

\section{Techniques of data analysis}
\label{techniques}

\subsection{Computation of line parameters}
\label{furhter}

Line parameters have been calculated for Line I and III from each
spectrogram of our time series as follows: 

The continuum intensity (hereafter $I_{\rm con}$) has been determined as the
average intensity value of 5 points, placed around the position of the maximum value of intensity in the right wings of Line I and Line III, respectively.

The position of the minimum of the lines has been determined by
fitting a fourth order polynomial to 12 points (Line I), 10 points
(Line III) and 8 points (terrestrial line), at the bottom of each line. By
taking the intensity value at this position, the line core
intensity of the solar lines (hereafter $I_{\rm I}$ and $I_{\rm III}$) is obtained.

To measure the line core velocities the following procedure has
been used. All spectra have been averaged in space and time, thus obtaining a mean profile on which the minimum positions of the two solar and the terrestrial lines have been computed. The absolute distance in wavelength between  Line I and the telluric line, $\overline{d_{\rm I}}$, and between Line III and the telluric line, $\overline{d_{\rm III}}$, has been adopted as the reference zero-velocity distance. This assumption is justified because we are only interested in relative velocity variations and not in absolute velocity values. The evaluation of the minimum positions has been performed also on each individual profile. By calculating the distances from Line I and Line III to the telluric
line, $d_{\rm I}$ and $d_{\rm III}$, the Doppler velocities of each line have 
been derived as the differences $\overline{d_{\rm I}}-d_{\rm I}$ and
$\overline{d_{\rm III}}-d_{\rm III}$. Instead of converting the Doppler
shifts into velocities hereafter the velocities will be referred in terms of
distances, that is, $V_{\rm I}=\overline{d_{\rm I}}-d_{\rm I}$ and $V_{\rm III}=\overline{d_{\rm III}}-d_{\rm III}$. 

Because of a blend in the upper left wing of Line III, the equivalent widths
of both solar lines (hereafter $EW_{\rm I}$ and $EW_{\rm III}$) have been
computed by integration of the area enclosed by each line profile below the 90\%-level of the continuum. 

Additionally, the full widths at half maximum for Line I and III (hereafter
$HW_{\rm I}$ and $HW_{\rm III}$) have been computed. 

In the following we will refer to $I$, $V$, $EW$, $HW$, (without subscripts)
when a common aspect in both lines is described. The values of the different line parameters (i.e. continuum intensity ($I_{\rm con}$), line core intensity ($I$), line core velocity ($V$), equivalent width ($EW$) and full width at half maximum ($HW$)) at each point $x$ along the spatial direction are arranged {\it vs.} time ($t$) in a 2D representation that from now on will be termed ``image'' (e.g. image of intensities, equivalent widths, etc.).

\subsection{Filtering of p-modes}
\label{pmodes}

The influence of p-modes in our images has been minimised by applying a
subsonic filter in the Fourier space ($k_x,\omega $). Only those spectral
components below the straight line $\omega =v_p k_x$ are retained whereas the
rest are set to zero (Deubner \cite{deubner88}). The constant $v_p$ defines
the cutoff, i.e. the maximum phase velocity admitted by the filter. Values of
$v_p$ ranging from 3 km\,s$^{-1}$ to 6 km\,s$^{-1}$ have been tested
obtaining similar results, and $v_p$ = 5 km\,s$^{-1}$ has been finally taken as a conservative choice.

Since the granular brightness fluctuation is a 2D spatial pattern, a 3D description, $g(x,y,t)$, would be desirable to perform a more complete spectral analysis --here $g$ means continuum intensity or any other spectroscopic parameter. However, our spatial information is restricted to only one dimension (along the spectrograph slit, $x$-$axis$, perpendicular to the dispersion direction), so that we have been compelled to work in 2D, ($x,t$), and consequently to obtain the Fourier transform $\overline G(k_x,\omega )$ which represents the integration of the 3D Fourier transform, $G(k_x,k_y,\omega )$, over $k_y$.

In effect, let $g(x,y,t)$ be the 3D array consisting of a time series of 2D images describing granulation parameters (intensities, Doppler velocities, equivalent widths, etc). The relation between $G$ and $\overline G$ is derived from the inverse Fourier transform formula for the case $y=0$,
i.e. $g(x,0,t)$:
\begin{eqnarray}
g(x,0,t)=\int\int \overline G(k_x,\omega)e^{2\pi i(k_x x + \omega t)}dk_x d{\omega}\,,
\label{equ3}
\end{eqnarray}
where
\begin{equation}
\overline G (k_x,\omega)=\int G(k_x,k_y,\omega)dk_y\,.
\label{equ4}
\end{equation}

According to Eq. \ref{equ4}, the information at a given $k_{x}$ is
contaminated by the contribution from spectral components lying onto a
parallel to the $k_{y}$-$axis$ at this specific $k_{x}$. This merging of
spectral information evidences that subsonic filtering would be desirable in
$G(k_x,k_y,\omega )$ instead of in $\overline G(k_x,\omega )$, the filter being defined by $\omega = v_p k$, where $k=\sqrt{k_x^2 + k_y^2}\,$ (see Title et al. \cite{title89}).

\begin{figure}[h]
\centering
\includegraphics[width=9cm,height=6cm]{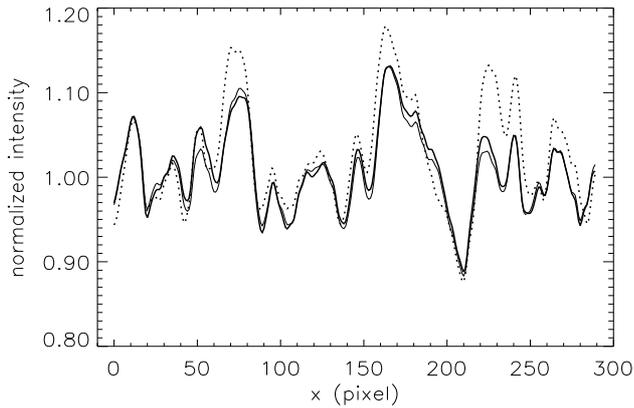}
\caption[]{Intensity profiles of solar granulation along the spatial coordinate $x$: Original (dotted) and after 2D (thick solid) and 3D (thin solid) subsonic filtering.}
\label {3D2D}
\end{figure}

For testing to what extent the 2D treatment is a reliable approximation in
the filtering process we have performed a numerical experiment by using an 80
min time series of 330 $\times$ 330 pixel high resolution images of
granulation taken at the SVST (Swedish Vacuum Solar Telescope, Observatorio
Roque de los Muchachos, La Palma), with a pixel size of $0\,\farcs062\times
0\,\farcs062$ and a time spacing of 18.9 s, see Hirzberger et
al. (\cite{hirzberger97}). A 2D array ($x,t$) has been extracted from the
original 3D array by fixing the $y$-coordinate in all granulation images;
the resulting array emulated the one obtained for the continuum from the
VTT's time series of spectrograms (that we have termed ``image'' of continuum
intensity). We also have averaged over 8 pixel in the $y$-direction to simulate the slit width of the spectrograph (cf. Sect. \ref{obs}). Then, 3D and 2D subsonic filtering with a common cutoff have been applied to the original 3D and to the extracted 2D array, respectively. We have repeated the 2D extraction at the same $y$-position but this time from the filtered 3D array, and compared with the results of the 2D filtering. The match of the results has improved when a gradual rather than an abrupt cutoff has been considered in the 2D filter. A cosine bell above the line $\omega = v_p k_x$, over the 10\% of the length of the $\omega$-$coordinate$ has been applied. Figure \ref{3D2D} shows a representative example of the results for the case of one scan in the $x$-direction. Note that the differences between the curves from 3D and 2D filtering -- thin and thick solid lines, respectively -- are small as compared with the original unfiltered curve -- dotted line.

\subsection {Calculation of coherence and phase difference spectra}

Analysis of coherence and phase differences between intensity ($i$) and
velocity ($v$) fluctuations of different spectral lines provides an adequate
diagnostic to discriminate between the various velocity fields that
contribute to the observed dynamics in the solar photosphere. Thus, [$v$--$v$],
[$i$--$i$] and [$v$--$i$] coherence and phase difference spectra are usually
calculated with respect, separately, to both wavenumber, $k_x$, and temporal
frequency, $\omega$ (see Deubner \& Fleck \cite{deubner89} and references therein).

Let $g_1(x,t)$ and $g_2(x,t)$ be 2D arrays representing the variation in
space and time of two granulation parameters. We have calculated
the coherence, $p_{12}$, and the phase difference, $\phi_{12}$, spectra
between these two arrays, with respect to $k_x$, as follows (see Edmonds \&
Webb \cite{edmonds72}; Deubner \cite{deubner88})
\begin{equation}
p_{12}(k_x)= {{\Big\vert \displaystyle\sum_t
    G_1(k_x,t)\,G_2^*(k_x,t)\Big\vert}\over
{\displaystyle\sum_t \Big\vert G_1(k_x,t) \, G_2^*(k_x,t)\Big\vert}}
\label{eq:ff4}
\end{equation}
\begin{equation}
\Phi_{12}(k_x) = {\rm arctan} \Biggl( {{{\rm Im}\,\displaystyle\sum_t
  G_1(k_x,t) \, G_2^*(k_x,t)}\over
{{\rm Re}\,\displaystyle\sum_t G_1(k_x,t) \, G_2^*(k_x,t)}}\Biggr)
\label{eq:ff5}
\end{equation}
where $^*$ stands for complex conjugate, $G_1$ and $G_2$ are the Fourier
transforms of $g_1$ and $g_2$ with respect to the spatial variable and the
summations are taken over times $t$. An interchange in the role of the spatial and temporal variables allows to obtain, similarly, coherence and phase difference spectra with respect to $\omega$.

\subsection {Heights of formation of the spectral lines}
\label{hof}
We aim at studying the variation of some physical parameters
with height in the photosphere. Therefore, it is appropriate to
determine how and where the two spectral lines respond to the temperature and
velocity field fluctuations. The response functions (RF) describe the sensitivity of the intensity profile to different photospheric quantities, like for instance temperature $T$ or line of sight velocity $V_{\rm LOS}$ (Ruiz Cobo \& del Toro Iniesta \cite{ruizcobo94}; S\'anchez Almeida et al. \cite{sanchez96}). We have calculated the RF of Line I and III to $T$ perturbations ($R_{T}$) and $V_{\rm LOS}$ perturbations ($R_{V_{\rm LOS}}$), applying the SIR code (Ruiz Cobo \& del Toro Iniesta \cite{ruizcobo92}). We refer also to Sheminova (\cite{sheminova98}), Gadun et al. (\cite{gadun97}, \cite{gadun99}) and P\'erez Rodr\'{\i}guez \& Kneer (\cite{perez2002}) for other calculations of response functions. \\ 

Line I is strong enough to consider that NLTE effects may cause significant
errors in the evaluation of the RF. We have used departure
coefficients $\beta_{\rm low}$ and $\beta_{\rm up}$ obtained by Shchukina \& Trujillo Bueno for the quiet sun model of Maltby et al. (\cite{maltby}, MACKKL model) and a 63 levels model atom of neutral iron. $\beta_{\rm low}$ ($\beta_{\rm up}$) stands for the ratio between the population of the lower (upper) atomic level evaluated in NLTE and LTE conditions, respectively. The absorption 
coefficient $K^{\rm NLTE}$ and the source function $S_{\nu}^{\rm NLTE}$ can be 
written as:
\begin{eqnarray}
K^{\rm NLTE} & = & 
\beta_{\rm low}K^{\rm LTE}\left(\frac{1-\beta_{\rm up}/\beta_{\rm low}e^{-h\nu/kT}}{1-e^{-h\nu/kT}}\right)\nonumber\\
 & \simeq & \beta_{\rm low}K^{\rm LTE}
\label{absorpt}
\end{eqnarray}
\begin{eqnarray}
S_{\nu}^{\rm NLTE} & = & \frac{2h\nu^{3}}{c^{2}} \frac{1}{\beta_{\rm low}/\beta_{\rm up}
e^{h\nu/kT}-1}\nonumber\\
 & \simeq & \frac{2h\nu^{3}}{c^{2}}
 \frac{\beta_{\rm up}}{\beta_{\rm low}}e^{-h\nu/kT} \simeq 
\frac{\beta_{\rm up}}{\beta_{\rm low}}S_{\nu}^{\rm LTE}
\label{source}
\end{eqnarray}

\begin{figure}[t]
\centering
\includegraphics[width=8cm,height=6cm]{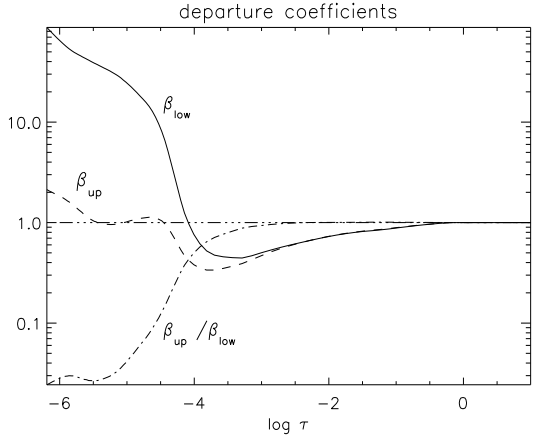}
\caption[]{Departure coefficients of Line I calculated for the quiet sun
model of Maltby et al. (\cite{maltby}, MACKKL model) and a \ion{Fe}{}--atomic 
model with 63 levels, where $\beta_{\rm low}$ and $\beta_{\rm up}$ are the departure coefficients of the lower and upper atomic level, respectively. The values have been kindly provided by Natasha Shchukina \& Javier Trujillo Bueno.}
\label{departure}
\end{figure}

The approximation in the second term of these equations is valid when we are 
able to neglect the stimulated emission ($h\nu >>  kT$), like in the case of 
solar lines in the visible range of the spectrum. In Fig. \ref{departure}, 
$\beta_{\rm low}$, $\beta_{\rm up}$ and its ratio is plotted versus $\tau$, the continuum optical depth at 5000 \AA. Using Eq. \ref{absorpt}, it is clear 
that the absorption coefficient is decreased by NLTE effects in the range 
of $\log \, \tau$ = 0 to $-$4 and increased from $\log \, \tau$ = $-$4 on, 
respectively, because of the behaviour of $\beta_{\rm low}$. The NLTE source 
function shows a strong decrease with respect to the Planck function from 
$\log \, \tau$ = $-$3 on caused by the behaviour of the ratio 
$\beta_{\rm up}/\beta_{\rm low}$ (see Eq. \ref{source}).

By substituting $K^{\rm NLTE}$ and $S_{\nu}^{\rm NLTE}$ for $K$ and $S_{\nu}$ in the expression of the response function (see Ruiz Cobo \& del Toro Iniesta \cite{ruizcobo94}) we have an suitable approximation of the RF which neglects changes in the departure coefficient but takes into account the more important NLTE effects (for a discussion see Socas-Navarro et al. \cite{socas}).\\

To evaluate the RFs we have used the MACKKL model, assuming a constant macroturbulence $\xi_{\rm mac}$ = 1.7 km\,s$^{-1}$, a constant microturbulence $\xi_{\rm mic}$ = 0.9 km\,s$^{-1}$, a collisional damping enhancement factor $E$ = 1.3, and a $\log \, gf$ value = $-$1.23 for Line I and values of $\xi_{\rm mac}$ = 1.3 km\,s$^{-1}$, $\xi_{\rm mic}$ = 0.9 km\,s$^{-1}$, $E$ = 2.9 and $\log \, gf$ = $-$0.66 for Line III, respectively. With these selection of parameters the synthesised profiles match the mean observed ones. For the calculation of the geometrical heights scale of this model we have solved the hydrostatic equilibrium equation considering a boundary condition for the surface gas pressure of 0.36 dyn/cm$^{2}$ (MACKKL model).\\

\begin{figure}[t]
\centering
\includegraphics[width=9cm,height=9cm]{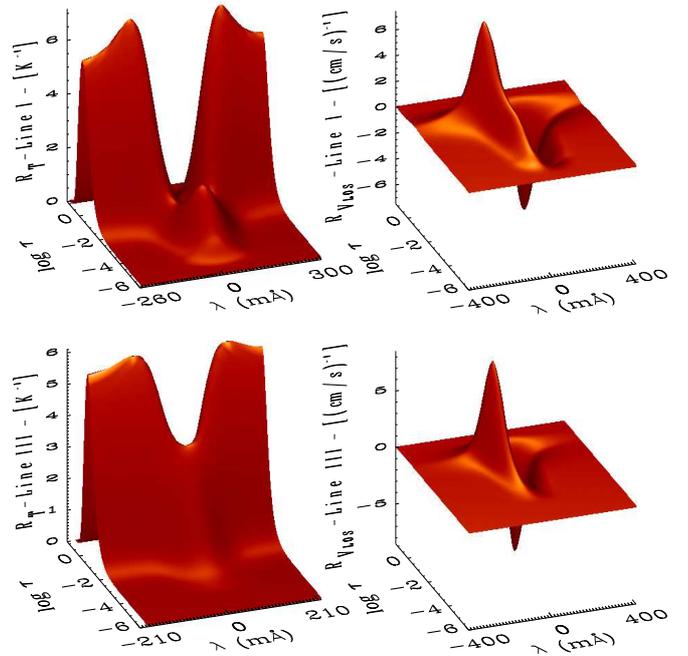}
\caption[]{Plot of the RF for $T$ (left panels) and $V_{\rm LOS}$ 
  perturbations (right panels) for Line I (upper panels) and III (lower
  panels), where $R_{T}$ and $R_{V_{\rm LOS}}$ are measuring the change of 
intensity for $T$ perturbation of 100 Kelvin and $V_{\rm LOS}$ perturbations of 1 km\,s$^{-1}$ at each layer of width $\Delta \log \, \tau$ = 0.1.}
\label{resp}
\end{figure}

In the left panels of Fig. \ref{resp} we can see the RFs to $T$ perturbations for Line I (top) and Line III (bottom) and in the right panels the corresponding response functions to $V_{\rm LOS}$ perturbations, respectively. The RFs plotted in this figure stand for the change of intensity at each wavelength when a local perturbation of 100 K and 1 km\,s$^{-1}$ (the order of  magnitude of $T$ and $V_{\rm LOS}$ perturbation we expect in granulation) is introduced at each layer in $\log \, \tau$. The two selected spectral lines are well differentiated in altitude and span a large height/optical depth range of the photosphere.\\

\begin{figure}
\centering
\includegraphics[width=9cm]{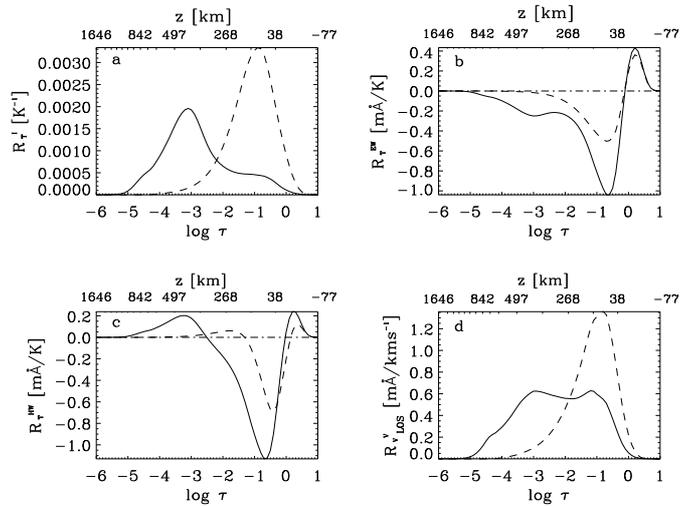}
\caption[]{Panels a, b and c correspond to the RF of $I$, $EW$ and $HW$ to a 
$T$ perturbation of 100 K. Panel d shows the RF of $V$ to $V_{\rm LOS}$ 
perturbation of 1km\,s$^{-1}$ at each layer of width $\Delta \log \, \tau$ = 
0.1. RFs of Line I and III are represented as solid and dashed lines, 
respectively.}
\label {rfresvel}
\end{figure}

Following Ruiz Cobo \& del Toro Iniesta (\cite{ruizcobo94}) the RFs of the line parameters $I$, $V$, $EW$ and $HW$ to $T$ and $V_{\rm LOS}$ perturbations can be straight forwardly expressed in terms of the RF of intensity. In Fig. \ref{rfresvel} we have plotted the RF of $I$, $EW$, $HW$ for $T$ perturbations and the RF of $V$ for $V_{\rm LOS}$ perturbations. Each line parameter is particularly sensitive to one photospheric quantity. However, all the different line parameters are sensitive to photospheric quantities in a broad range of the photosphere.

Especially difficult is the case of the RF of $V_{\rm I}$, which shows a maximum sensitivity in the range spanning from $\log \, \tau$ $\simeq$ $-$1 to $\log \, \tau$ $\simeq$ $-$3.1, so that we are compelled to work with sampled information originating from very different heights of the photosphere.

The RF of $EW$ is mainly influenced by the 
effect that in higher layers (above $\log \, \tau$ = 0) an increasing 
temperature yields a negative lobe in the RF caused by a smaller absorption 
due to the decrease of \ion{Fe}{i} abundance (Saha formula), while in lower layers (below $\log \, \tau$ = 0) a positive lobe in the RF is caused by an 
increase of continuum intensity. 

The RF of $HW$ is influenced by the same 
effect like in the case of $EW$ plus an additional broadening caused by the 
increase of Doppler width and resulting in a second positive lobe in the 
response function in higher layers. Note, that the height of formation for 
$EW$ and $HW$ would change, if we would use other definitions like the centre 
of gravity of the response function instead of its peak. 

The sensitivity of $I$, $EW$ and $HW$ to $V_{\rm LOS}$ perturbations and of $V$
to $T$ perturbations is negligible, because our model did not represent strong velocity gradients.\\

\begin{table}
\begin{flushleft}
\tiny
\begin{tabular}{lllcccc}
\hline
Line&[$\log \, \tau$]&&${I}$&$EW$&$HW$&$V$ \\
\hline
I&range&\vline&[0., $-$5.]&[1., $-$5.]&[0.5, $-$5.]&[0.,
$-$5.] \\
&HOF&\vline&$-$3.1&$-$0.6&$-$0.7&[$-$1.2, $-$3.1] \\
\hline
III&range&\vline&[0., $-$3.]&[1., $-$3.]&[1., $-$3.]&[0.,
$-$3.] \\
&HOF&\vline&$-$0.9 & $-$0.6 & $-$0.4 &
$-$0.8 \\ 
\hline
\end{tabular}
\end{flushleft}
\normalsize
\caption[]{Column 3, 4, 5, 6: total range of sensitivity to $T$ and $V_{\rm
LOS}$ perturbations and height of formation (HOF) of $I$, $EW$, $HW$ and 
$V$ in optical depth, for Line I and III, respectively.}
\label{maxima1}
\end{table}
\begin{table}
\tiny
\begin{flushleft}
\begin{tabular}{lllcccc}
\hline
Line&[km]&&$I$&$EW$&$HW$&$V$ \\
\hline
I&range&\vline&[0, 1065]&[$-$66, 1065]&[$-$36, 1065]&[0, 1065] \\
&HOF&\vline&435&75&90&[166, 435] \\
\hline
III&range&\vline&[0, 421]&[$-$66, 421]&[$-$66, 421]&[0, 421] \\
&HOF&\vline&120&75&46&105 \\
\hline
\end{tabular}
\end{flushleft}
\normalsize
\caption[]{Column 3, 4, 5, 6: total range of sensitivity to $T$ and
$V_{\rm LOS}$ perturbations and height of formation (HOF) of $I$, $EW$, 
$HW$ and $V$ in geometrical height, for Line I and III, respectively.}
\label{maxima2}
\end{table}

In Tables \ref{maxima1} and \ref{maxima2} we have summarised the results
regarding the height of formation. Total ranges of the sensitivity of different line parameters to $T$ or $V_{\rm LOS}$ perturbations and assumed heights of formation are given in optical depth and geometrical height.\\ 

The formation height concept is somewhat inexact and to ascribe certain physical properties derived from spectral lines to a given height in the atmosphere is maybe misleading. However, it is useful for interpreting the results qualitatively. In subsequent papers the observed data will be inverted by means of using
the complete response functions to the different physical quantities. This
will allow us to obtain results free from the simplification assumed here.

\subsection{Selection of granular and intergranular regions.}

\begin{figure*}[t]
\centering
\includegraphics[width=12cm,height=6cm]{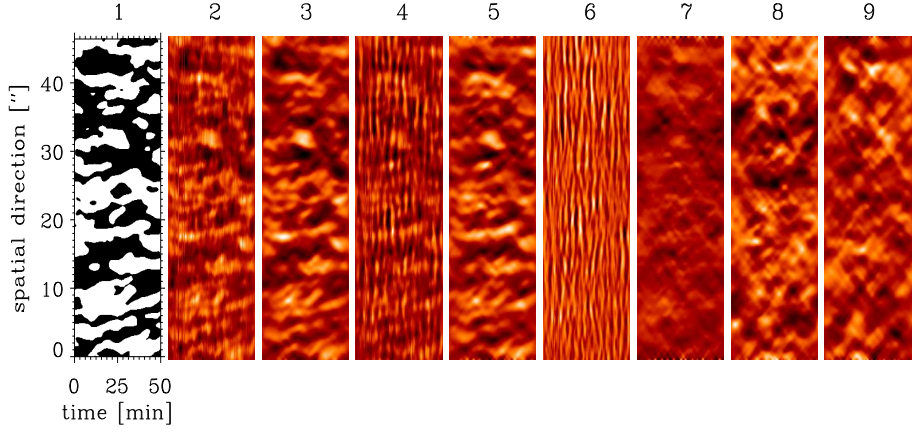}
\caption[]{Images representing the variation of continuum brightness,
  velocities and line core intensities for Line I and Line III. From the left
  to the right: (1) $I_{\rm con}$ binary image, (2) $I_{\rm con}$ unfiltered,
  (3) $I_{\rm con}$ filtered from p-modes, (4) $V_{\rm III}$ unfiltered, (5)
  $V_{\rm III}$ filtered, (6) $V_{\rm I}$ unfiltered, (7) $V_{\rm I}$
  filtered, (8) $I_{\rm III}$ filtered, (9) $I_{\rm I}$ filtered. Horizontal
  $axis$ is time and vertical $axis$ space. To facilitate a visual comparison between velocity and intensity images, we have reversed the scale of velocity values so that granular velocities (negative) are represented also as bright areas. See the text for a detailed description of the different panels.}
\label{gdisp}
\end{figure*}

We must first consider how to separate granules from
intergranular lanes. In general we can formulate the problem in this way: granules (intergranular lanes) are defined by all those pixels (i, j) with intensity and velocity values satisfying the following condition:\\ 

\noindent{\it Granules:}

\begin{equation}
I_{\rm ij}>(1+q) \overline {I} \hspace{0.4cm}; \hspace{0.4cm}  V_{\rm ij}>(1+\hat q) \overline {V}
\label{gran}
\end{equation}

\noindent{\it Intergranular lanes:}

\begin{equation}
I_{\rm ij}<(1-q) \overline {I} \hspace{0.4cm}; \hspace{0.4cm}  V_{\rm ij}<(1-\hat q) \overline {V}
\label{ingran}
\end{equation}

where $I_{\rm ij}$ and $V_{\rm ij}$ are intensity and velocity values at each pixel in space and time, and $\overline {I}$, $\overline {V}$ the mean values of
brightness and velocity over the whole image, respectively. $q$ and $\hat q$ are threshold parameters defining a granular--intergranular transition region.

Most of the authors have only used an intensity criterion by assuming $q=0$. Keil (\cite{keil}) defined granules (intergranular lanes) as those regions where the difference between the local continuum intensity and the mean continuum intensity ($\overline {I}$) is above (below) 8 $\%$ ($q = 0.08$). This percentage was reduced to 1$\%$ by Muller \& Roudier (\cite{muller84}) and to 4 $\%$ by Karpinsky (\cite{karpinsky}).

The granulation as a convective phenomenon is better detected in the deepest observable layers of the photosphere. The continuum intensity reflects the temperature fluctuations in these layers, whereas line core velocities of Line III and Line I stem from higher layers. Since we do not have information about velocities from the deepest layers it seems more reasonable to use the single criterion based on intensity. Thus, finally this single criterion has been adopted, assuming $q$=0.

\section {Global correlations}
\label{global}
\subsection {A glance to the images}
\label{glance}

Figure \ref{gdisp}, composed of nine vertical panels, summarises some of the
results of our VTT spectrogram series. Each image, composed of 152 columns,
represents the variation of a spectroscopic parameter along the spectrograph
slit (vertical), at successive times throughout the series (horizontal). The contents of the individual panels or images are as follows:

\begin {itemize}
\item[$\bullet$] 1: binary image of the continuum brightness showing the time evolution of the granular (white) and intergranular (black) regions along the spectrograph slit.
\item[$\bullet$] 2 and 3: unfiltered and filtered continuum brightness
($I_{\rm con}$). The very fine vertical structures (high temporal frequency)
observed in our raw images, caused by differential seeing distortion between
successive spectrograms or by a residual tracking error are removed by the subsonic filter.
\item[$\bullet$] 4 and 5: unfiltered and filtered line core velocities for Line III ($V_{\rm III}$).
\item[$\bullet$] 6 and 7: unfiltered and filtered line core velocities for
  Line I ($V_{\rm I}$).
\item[$\bullet$] 8 and 9: filtered line core intensities for Line III and I ($I_{\rm III}$, $I_{\rm I}$), respectively.
\end {itemize}

From these images we can extract the following qualitative information:

\begin{table*}[t]
\centering
\scriptsize
\begin {tabular}{lllrrrrrrrr}
\hline
&&\vline&$V_{\rm III}$&$V_{\rm I}$&$I_{\rm III}$&$I_{\rm I}$&$EW_{\rm III}$&$EW_{\rm I}$&$HW_{\rm III}$&$HW_{\rm I}$ \\

\hline
$I_{\rm con}$&unfiltered&\vline&$-$0.61&$-$0.17&0.14&$-$0.22&0.50&0.34&$-$0.43&$-$0.37 \\
&filtered&\vline&$-$0.87&$-$0.49&0.03&$-$0.44&0.69&0.58&$-$0.46&$-$0.37 \\
\hline
$V_{\rm III}$&unfiltered&\vline&&0.74&0.02&&$-$0.48&&0.25 \\
& filtered&\vline&&0.62&0.04&&$-$0.79&&0.30 \\
\hline
$V_{\rm I}$&unfiltered&\vline&&&&0.07&&$-$0.18&&0.02 \\
& filtered&\vline&&&&0.48&&$-$0.51&&0.24 \\
\hline
$I_{\rm III}$&unfiltered&\vline&&&&0.32&$-$0.54&&0.63 \\
&filtered&\vline&&&&0.24&$-$0.44&&0.68 \\
\hline
$I_{\rm I}$&unfiltered&\vline&&&&&&$-$0.83&&0.56 \\
&filtered&\vline&&&&&&$-$0.78&&0.56 \\
\hline
$EW_{\rm III}$&unfiltered&\vline&&&&&&0.79&$-$0.31 \\
&filtered&\vline&&&&&&0.82&$-$0.33 \\
\hline
$EW_{\rm I}$&unfiltered&\vline&&&&&&&&$-$0.09 \\
&filtered&\vline&&&&&&&&$-$0.03  \\
\hline
$HW_{\rm III}$&unfiltered&\vline&&&&&&&&0.45 \\
&filtered&\vline&&&&&&&&0.54  \\
\hline
\end {tabular}
\normalsize

\vspace{0.2cm}
\caption []{Correlation coefficients between the fluctuations of different line parameters for filtered and unfiltered data.} 
\label{ccc}
\end{table*}

\begin {itemize}
\item[$\bullet$] The filtering method has apparently removed the fluctuations associated with the p-modes.
\item[$\bullet$] The expected difference between granules and intergranular
  lanes can be clearly seen. The width of the structures is also the one
  expected (cf. Bray, Loughhead \& Durrant \cite{bray}). There exist granules
  and intergranular lanes that persist over the full observing period in agreement with Sobotka et al. (\cite{sobotka}).
\item[$\bullet$] In the different images we see small-scale structures that
  could correspond to fine features inside the granules or perhaps to non-convective motions present in the upper layers.
\item[$\bullet$] As derived from numerical experiments we have done, the small amplitude cross-shaped structure visible in images 7, 8 and 9 is caused by the cutoff of the p-modes filter, even in the case of smoothing the transition from 1 to 0, within a narrow band close to the cutoff.
\item[$\bullet$] The granules suffer different kinds of evolutionary events like expansions, merging and fragmentation as it is observed in 2D images (e.g.
Hirzberger et al. \cite{hirzberger99}).
\end{itemize}

\subsection {Tentative explanation of global correlations}
The filtering process allows us to discriminate between two different
domains, namely: a) the oscillatory component, representing the p-modes
together with residual motions induced by the tracking error and/or
seeing, and b) the convective component, representing the signature of the
convective overshooting in the photosphere together with the gravity waves
(Deubner \& Fleck \cite{deubner89}; Bonet et al. \cite{bonet91}; Komm et
al. \cite{komm91c}). We have calculated the global correlations between
different line parameters in order to get some preliminary hints about how
the dynamics at different heights change through the photosphere and we have
summarised these results in Table \ref{ccc}. In most of the cases the
filtering of p-modes in our images has increased the degree of correlation. In Sect. \ref{spat} we will enter in more detail by performing a coherence and phase shift spatial analysis to obtain information about the penetration of
different structures across the photosphere.\\

In the following we will remark the most significant results of correlation
coefficients presented in Table \ref{ccc} and give a tentative explanation in
terms of the RF plotted in Fig. \ref{rfresvel}:

\begin{figure}[]
\centering
\includegraphics[width=8cm,height=8cm]{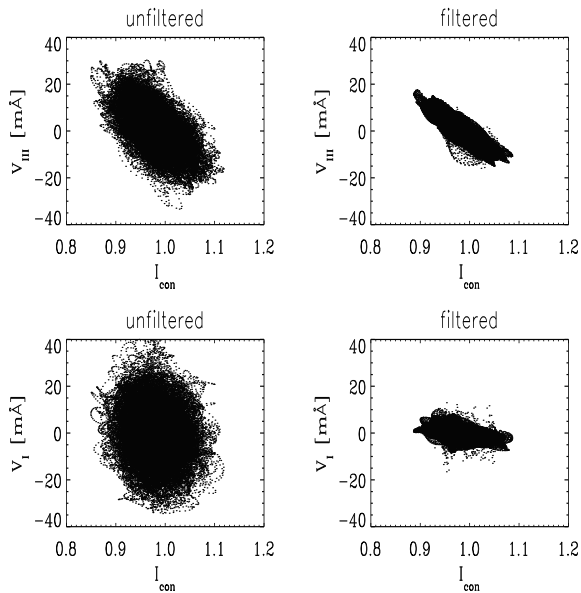}
\caption[]{Scatter diagrams of continuum intensity and velocity as measured
from Line III (upper panels) and Line I (lower panels) before (left panels) and
after filtering of p-modes (right panels).}
\label{coriv}
\end{figure}

\begin{itemize}
\item[$\bullet$] Correlations with velocities:\\
In Fig. \ref{coriv} we present a scatter plot of
$I_{\rm con}$ {\it vs.} $V$, before and after filtering the p-modes.
The high degree of correlation between $I_{\rm con}$ and $V_{\rm III}$ after
filtering the p-modes is a clear consequence of the convective nature of
granulation. The correlation between $I_{\rm con}$ and $V_{\rm I}$ is weaker
, thus reflecting a less efficient penetration of the convective velocity field into higher layers (Deubner \cite{deubner88}; Espagnet et al. \cite{espagnet95}). This behaviour is also confirmed by a decrease of the coherence between [$I_{\rm con} - V_{\rm I}$] with respect to that of [$I_{\rm con} - V_{\rm III}$] (see Fig. \ref{coh1}). Remember, that from the calculation of response
functions in Sect. \ref{techniques}, the height of formation (HOF) of
$V_{\rm III}$ can be assigned to $\log \, \tau$ = $-$0.8, while $V_{\rm I}$
originates from a broad range in optical depth [$-1.2$; $-3.1$], clearly higher in any case. Thus we are able to conclude that convective motions penetrate up to the highest layers of the photosphere, where the velocity pattern is still connected with the brightness pattern at the continuum layer.

The correlation between velocities at the two levels ($V_{\rm III}$ {\it vs.} $V_{\rm I}$) is high but decreases considerably when we
study only the convective component. This means, that the contribution of
oscillatory velocities due to the p-modes is significant at the height of
formation of both lines.\\
\item[$\bullet$] Correlations with line core intensity:\\
We find a weak negative correlation between the pair [$I_{\rm
  con}$ -- $I_{\rm I}$] due to the reversal of
temperature contrast in high photospheric layers (e.g. Deubner \cite{deubner88}; Collados et al. \cite{collados}; Rodr\'{\i}guez
Hidalgo et al. \cite{rodriguezhidalgo}). The inversion of temperature
contrast was predicted by theoretical models (Steffen et
al. \cite{steffen89}; Stein \& Nordlund \cite{stein89}; Gadun et al. \cite{gadun97}, \cite{gadun99}). It also explains the weak positive correlation between the pair [$V_{\rm I}$ -- $I_{\rm I}$].

We do not find any significant correlation between the pairs [$I_{\rm con}$ --
$I_{\rm III}$], [$V_{\rm III}$ -- $I_{\rm III}$] and [$I_{\rm I}$ -- $I_{\rm
  III}$]. This could be caused by the fact that the peak of the temperature
response function of $I_{\rm III}$ is placed $\sim$ $\log \, \tau$ = $-$0.9
(panel a of Fig. \ref{rfresvel}) just at the layer where the inversion of
temperature occurs: for the case of a granule, higher temperatures in
deeper layers compensate the variation of $I$ introduced by lower temperatures
in the layers above.

Berrilli et al. (\cite{berilli}), by studying two lines with formation heights about 68 km and 370 km, found correlations between line core velocity and line core intensity of opposite sign (positive for the line whose core is formed at $\sim 370$ km). It is well known that the granular brightness fluctuations decrease rapidly with increasing height, vanish at $\sim$140 km where the
inversion of temperature contrast takes place, and increase slightly
above this height again (Rodr\'{\i}guez Hidalgo et al. \cite{rodriguezhidalgo}). If we take into account that $I_{\rm III}$ and $I_{\rm I}$ can be ascribed to a height of $\sim$120 km and $\sim$435 km (see Table \ref{maxima2}), respectively, our results agree with the results obtained by Berrilli et al. (\cite{berilli}). Note, that our results are also in agreement with other works in the literature but there exist discrepancies concerning the height where the brightness fluctuations vanish (see Introduction).\\
\item[$\bullet$] Correlations with equivalent width:\\
For the pairs [$I_{\rm con} - EW_{\rm I}$] and [$I_{\rm con} - EW_{\rm III}$]
we find for the convective component significant positive correlation. From
panel b of Fig. \ref{rfresvel} we can deduce that granular line profiles have 
larger equivalent width than the intergranular ones, because granules are
much hotter below $\log \, \tau$ = 0 (where the RFs for both lines are
positive) and slightly cooler above $\log \, \tau$$ = -1$ (negative RFs).
Also Hanslmeier et al. (\cite{hanslmeier93}) reported a positive correlation
($\sim 0.53$) between these parameters studying four lines, having a line
core formation height between 70 and 150 km, although their data were not
filtered. The positive correlation has been explained (e.g. Gadun et
al. 1997) as due to the influence of oscillatory motions. Our data show the
opposite behaviour since we find enhanced correlation after filtering.

The negative correlation between the pairs [$V_{\rm III} - EW_{\rm III}$] and
[$V_{\rm I} - EW_{\rm I}$] can also be explained taking into account the
behaviour of equivalent width as stated in the previous paragraph.

For the pairs [$I_{\rm III} - EW_{\rm III}$] and [$I_{\rm I} - EW_{\rm I}$]
we have found for the convective component a high negative correlation for
Line I and a weak negative correlation for Line III. This can be explained
taking into account the behaviour of line core intensity that we comment in
case of correlation between $I_{\rm con}$ and $I$.

The equivalent widths of both lines are highly correlated. This can be explained by the fact that the temperature response functions for the two lines are very similar to each other (panel b of Fig. \ref{rfresvel}). An increase in temperature produce a similar change in equivalent width resulting in a strong correlation for the pair [$EW_{\rm III} - EW_{\rm I}$].\\

\item[$\bullet$] Correlations with full width at half maximum:\\
The RF of $HW$ for Line I has a strong negative lobe around $\log \, \tau$ = $-$1. As temperature fluctuations at these layers do not affect the continuum intensity, no significant correlation between $I_{\rm con} - HW$ can be expected. For Line III, a weak correlation between these parameters is observed, because the negative lobe of the corresponding response function is placed at deeper layers. Besides, a weak correlation for the pair [$HW_{\rm III} - HW_{\rm I}$] can be expected, as the response functions of both lines partially coincide.

Another case is the observed correlation between line core intensity
and half width. Turbulence and strong velocity gradients produce an enhancement of line width together with an increase of line core intensity, while a growing of line width induced by temperature changes always goes with a decrease of line core intensity. Thus, the significant correlation between $I$ and $HW$ indicates that either turbulence or strong velocity gradients must be important agents to explain the enhancement of the width of both lines. Nesis et al. (\cite{nesis99}) related $HW$ with an unresolved turbulent velocity field placed mainly at the granular borders.
\end{itemize}
 We have not found any significant values for the remaining correlations between the several line parameters.

\subsection {Line parameters in granular and intergranular regions}
Mean values of the relevant line parameters were calculated separately for
granular and intergranular regions (Table \ref{blp}).\\
\begin {table}[h]
\begin {tabular}{lrr}
Mean values  & Granular regions & Intergranular Lanes \\
\hline
${\rm I_{I}}$ & $0.26 \pm 0.01$ & $0.27 \pm 0.02$ \\
${\rm I_{III}}$ & $0.50 \pm 0.02$ & $0.50 \pm 0.02$ \\
${\rm EW_{I}}$ & $116.1 \pm 2.8$ & $112.7 \pm 3.5$\\
${\rm EW_{III}}$ & $41.3 \pm 1.2$  & $39.4 \pm 1.5$\\
${\rm HW_{I}}$ & $186.6 \pm 3.6$ & $189.2 \pm 4.6$ \\
${\rm HW_{III}}$ & $116.8 \pm 4.4$ & $120.6 \pm 4.8$\\
\hline
\end {tabular}
\caption []{Mean filtered values (with standard deviations) of line
  parameters in granular and intergranular regions. Granular (intergranular)
  regions are defined as those where $I_{\rm con}$ is above (below) its mean value.}
\label{blp}
\end {table}
As expected from the analysis of the correlation between the several line parameters, we do not find significant difference between the mean values of line core intensities above granules and intergranular lanes in case of Line III, since the information contained in $I_{\rm III}$ originates mainly from layers around the inversion of temperature where the brightness contrast between granules and intergranular lanes remains near 0. For Line I we find a slightly lower mean value of line core intensity above granules than above intergranular lanes, thus we can conclude that in higher layers granules appear slightly cooler than the surrounding intergranular lanes due to the inverted temperature contrast. This is in agreement with models of Steffen et al. (\cite{steffen89}) and Stein \& Nordlund (\cite{stein89}), which predict that temperature fluctuations should change sign in the overshooting layer, which should be cooler above the ascending granular parts and 'relatively' warmer above the descending intergranular regions.

For both lines we find enhanced values of $EW$ above granules. Gadun
et al. (1997) predict an increase of $EW$ in granules as observed. $HW$ shows higher values above the intergranular lanes, which has been also reported by
Nesis et al. (\cite{nesis90}) and Hanslmeier et al.(\cite{hanslmeier93}). 

\section {Spatial variations}
\label{spat}
\subsection {Coherence and phase shifts spatial spectra}
\begin{figure}
\centering
\includegraphics[width=8cm,height=3cm]{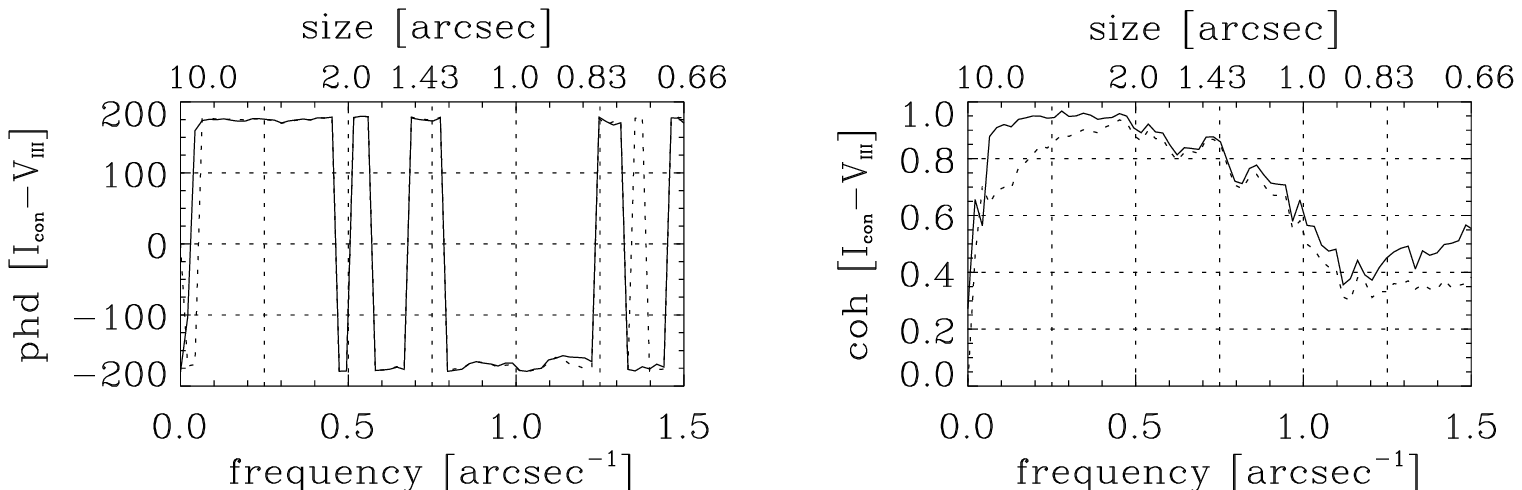}
\includegraphics[width=8cm,height=3cm]{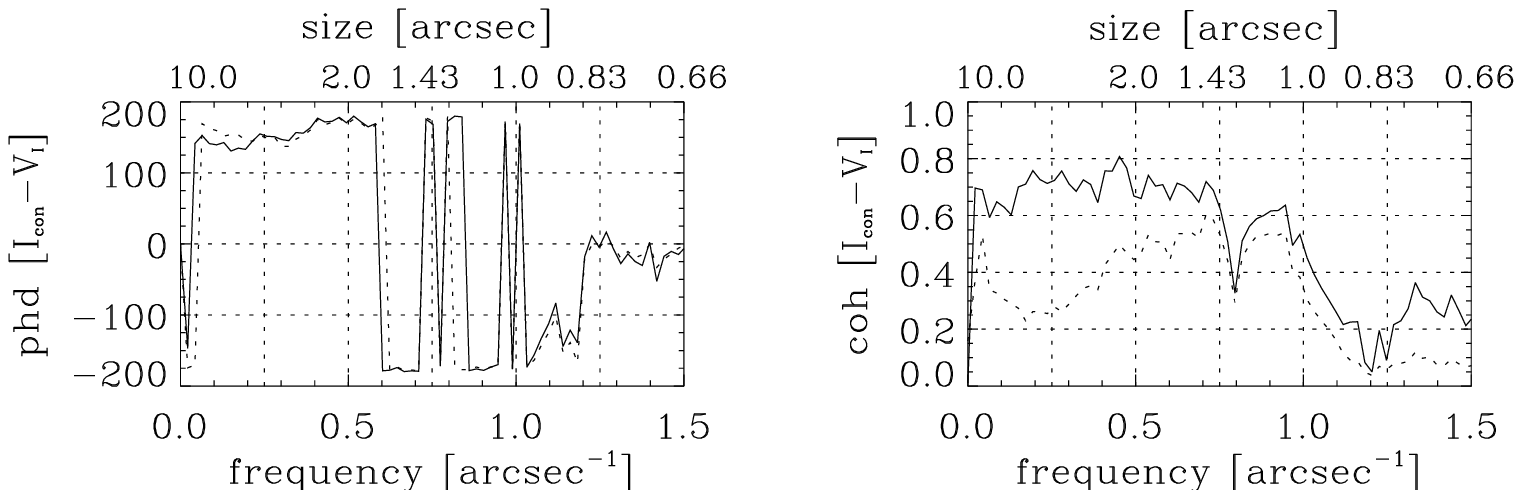}
\includegraphics[width=8cm,height=3cm]{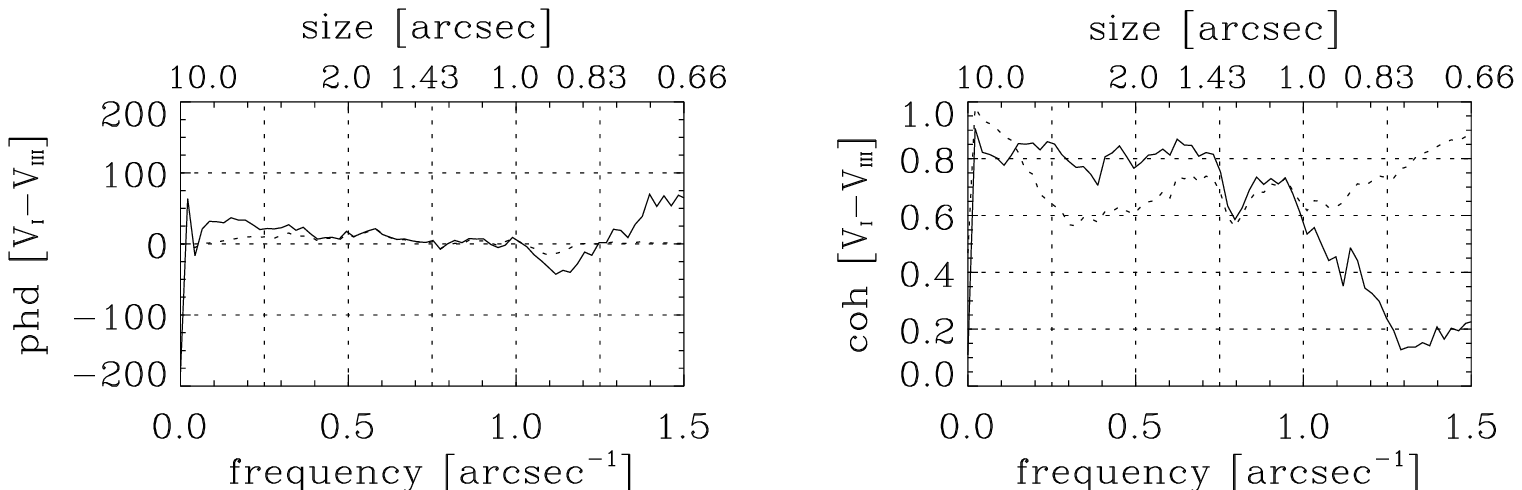}
\includegraphics[width=8cm,height=3cm]{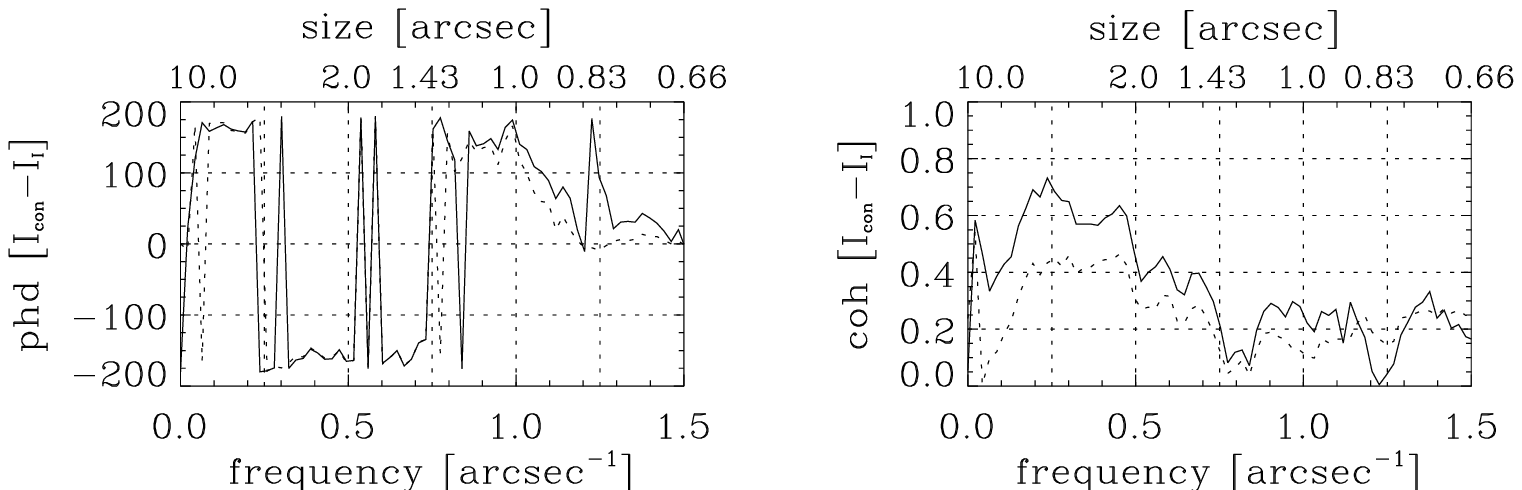}
\includegraphics[width=8cm,height=3cm]{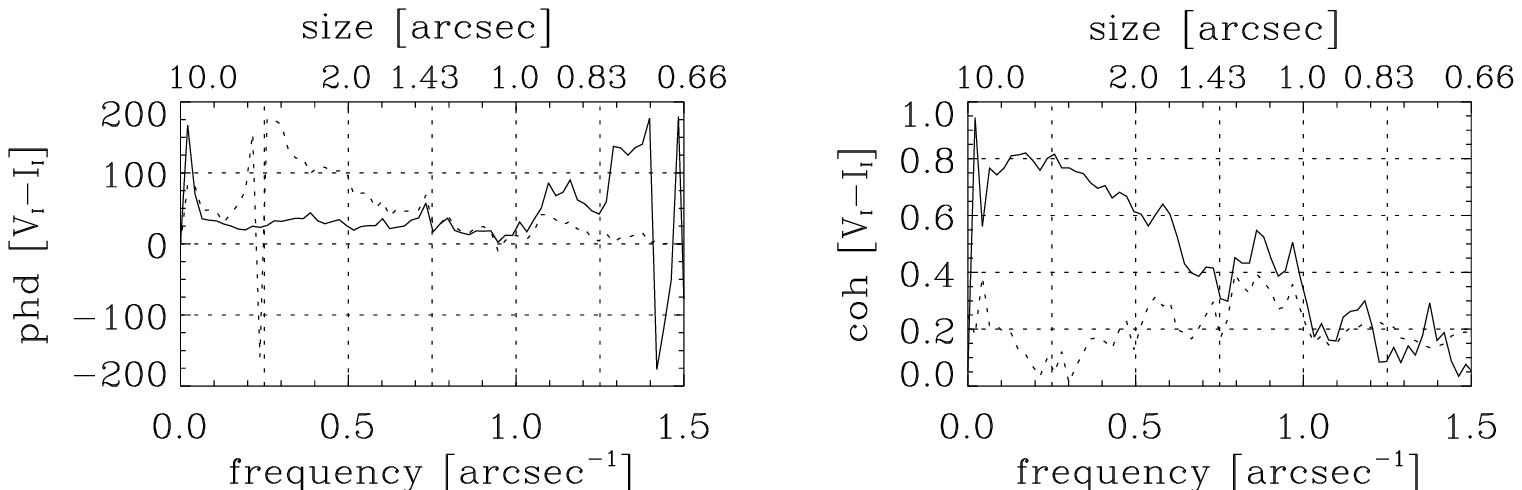}
\caption[]{Spatial variations : Average behaviour
of the coherence (coh) and phase shifts (phd) between the pairs [$I_{\rm con} -
V_{\rm III}$], [$I_{\rm con} - V_{\rm I}$], [$V_{\rm I} - V_{\rm III}$],
[$I_{\rm con} - I_{\rm I}$] and [$V_{\rm I} - I_{\rm I}$], before (dashed-dotted line) and after filtering (solid line) the p-modes.}
\label{coh1}
\end{figure}

An analysis of coherence and phase difference spectra between two photospheric 
lines formed at well separated heights gives information about how the
granulation phenomenon is related at these two layers (e.g. penetration of
granulation through the photosphere). In Fig. \ref{coh1}, the coherence and phase difference spectra between several line parameters are plotted. Only phase shifts with a corresponding coherence above $\sim$ 0.5 will be considered as reliable enough. Coherence and phase shift are represented {\it vs.} spatial frequency and result from temporal averages. In the present discussion we are assuming the upper limit of the granular scales at $\approx$ 4\,\farcs0.

First we want to remark that the computed coherence spectra increase
significantly in case of the filtered images due to the elimination of
p-modes. This is especially noticeable in case of Line I, because the
amplitude of p-modes increases with height in the photosphere.

From Fig. \ref{coh1} we can infer the following ranges of spatial scales
with significant coherence (i.e. $>$ 0.5): a) scales above $\sim$ 1\,\farcs0
for [$I_{\rm con} - V_{\rm I}$], [$I_{\rm con} - V_{\rm III}$], and [$V_{\rm
  I} - V_{\rm III}$]; b) scales between $\sim$ 9\,\farcs0 and 2\,\farcs1 for
[$I_{\rm con} - I_{\rm I}$]; and c) scales $\geq$ 1\,\farcs6 for [$V_{\rm I} -
I_{\rm I}$]. These high values of coherence reveal that the granular
phenomenon at larger scales penetrates up to the highest levels of the
photosphere (at least up to $\sim$ 435 km). Lets study more deeply these
three cases.

\begin{itemize}
\item[$\bullet$] [$I_{\rm con} - V_{\rm I}$], [$I_{\rm con} - V_{\rm III}$], and [$V_{\rm I} - V_{\rm III}$]:\\
$I_{\rm con}$ leads $V_{\rm III}$ by 180$^{0}$ and $V_{\rm I}$ by a decreasing
 amount from 180$^{0}$ to $\sim$ 150$^{0}$ in the granular range with high coherence values (1\,\farcs0 -- 4\,\farcs0). For larger scales ($>$ 4\,\farcs0) the phase difference between $I_{\rm con}$ and $V_{\rm I}$ continues decreasing. Consequently with these results  $V_{\rm I}$ leads $V_{\rm III}$ by an amount augmenting from 0$^{0}$ to $\sim$ 20$^{0}$ in the same granular
range. Deubner (\cite{deubner89a}) finds similar results by studying the
lines \ion{C}{i} 5380 and \ion{Fe}{i} 5383, with corresponding heights of
formation $\sim$ 0 and 200 km above $\tau_{5000}$ = 1. Similarly to our case, 
he finds in the granular regime the \ion{Fe}{i} velocity signal leading the
\ion{C}{i} velocity by $10^{0}$ to $15^{0}$; however at the larger scales his
velocities are well in phase.\\

\item[$\bullet$] [$I_{\rm con} - I_{\rm I}$]:\\
The coherence at very large scale is below 0.5, then rises above 0.7 at about
5\,\farcs0 and drops abruptly at $\sim$ 2\,\farcs1. Thus, regarding the
brightness structure only the largest granular scales ($\sim$ 2\,\farcs0 --
4\,\farcs0) are detectable at highest photospheric levels. Nevertheless, the
brightness pattern at these layers is inverted with respect to that of the continuum as a consequence of the temperature inversion at $\sim$ 140 km.\\

\item[$\bullet$] [$V_{\rm I} - I_{\rm I}$]:\\
At the highest levels the brightness pattern is similar to that of velocities 
but only at the largest granular scales ($\sim$ 1\,\farcs6 --
4\,\farcs0). From the phase shifts measured for [$I_{\rm con} - I_{\rm I}$] and
[$I_{\rm con} - V_{\rm I}$], we would expect $V_{\rm I}$ leading $I_{\rm I}$
by an increasing value from $0^{0}$ at 1\,\farcs6 to $\sim$ $20^{0}$ at
4\,\farcs0. For \ion{Fe}{i} 5383, Deubner (\cite{deubner89a}) finds the line
core intensity leading the velocity in a similar amount, although with an
increasing trend toward mesogranular scales which can also be detected in the
very low frequency range of our figure. Deubner (\cite{deubner89a}) explains
this effect as well as the phase difference similar to that one that we have
encountered between $V_{\rm III}$ and $V_{\rm I}$ as caused by the increase
of thermal relaxation time with height and also by the presence of gravity waves superposed on the convective motions.
\end{itemize}

We have not found any coherence for the pairs [$I_{\rm con} - I_{\rm III}]$, [$V_{\rm III} - I_{\rm  III}$] and [$I_{\rm I} - I_{\rm  III}$] due to the small brightness contrast at layers close to the inversion of temperature, resulting in a decrease of the signal to noise ratio and a weakening of coherence.\\

Our results complement those by Deubner (\cite{deubner89a}) in the sense that 
we report information at complementary heights in the solar
photosphere. Espagnet et al. (\cite{espagnet95}) perform a phase and
coherence spectral analysis between intensity and velocity fluctuations as
measured at several positions (wavelengths) throughout the profile of the
\ion{Na$\rm D_{2}$}{} line ($\lambda$ 5890) which correspond to various heights in the photosphere ranging from the continuum level up to $\sim$ 550 km. 
In contrast with our findings they conclude that temperature fluctuations of granulation do not penetrate higher than about 60-90 km. By comparing the intensity fluctuations at the continuum level with those at different levels in the photosphere, they find high coherence for large granules ($>$1\farcs4) up to 60 km, vanishing at 90 km. For Nesis et al. (\cite{nesis88}) and Komm et al. (\cite{komm90}) the coherence of granules $>$ 1\,\farcs4 vanishes at 170 km above the continuum level. 

Regarding velocity features, Espagnet et al. (\cite{espagnet95}) compare the velocities at different layers with both, the continuum intensity and the velocity at the height of 30 km. From these coherence and phase shift analyses they conclude that the overshooting velocities of granules $>$ 1\,\farcs4 and $>$ 1\,\farcs6, respectively, cross the whole photosphere up to a level of at least 550 km, and that the height of penetration decreases with decreasing size. These results are in a closer agreement with ours except for two aspects:
1) they find almost perfect antiphase at all photospheric layers, whereas we
obtain that $I_{\rm con}$ leads $V_{\rm I}$ by a varying angle between
$180^{0}$ and  $\sim 150^{0}$ (this decreasing trend in the phase difference for large granules could be a signature of gravity waves in the upper photosphere); and 2) and velocities of granules larger than 1\,\farcs0 instead of 1\,\farcs4 penetrate into high layers (435 km).

\subsection {Dependence of line parameters with the distance to the granular border}
\label{dep}
In this section we intend to figure out how the line parameters vary as a
function of the normalised distance to the granular border. For this purpose
granules and intergranular lanes have been identified in the spatial direction 
(columns in the image of the continuum intensity) defining the granular
borders at the inflection points of the intensity profiles. The middle
position of every interval between two consecutive inflection points is
taken as the centre of the corresponding granule or intergranular lane.

For each granule the distances from the border to the centre
have been computed, normalised to unity and interpolated to the same number of distance steps, so that, independently of the granular size, the distance is zero at the border and 1 at the centre and containing 100 values in between. Similarly, for intergranular lanes the distances have been set to vary from zero at the border to $-$1 at the centre, independently also of
the intergranular lane size. 

\begin{figure}[h]
\centering
\includegraphics[width=4cm,height=3.0cm]{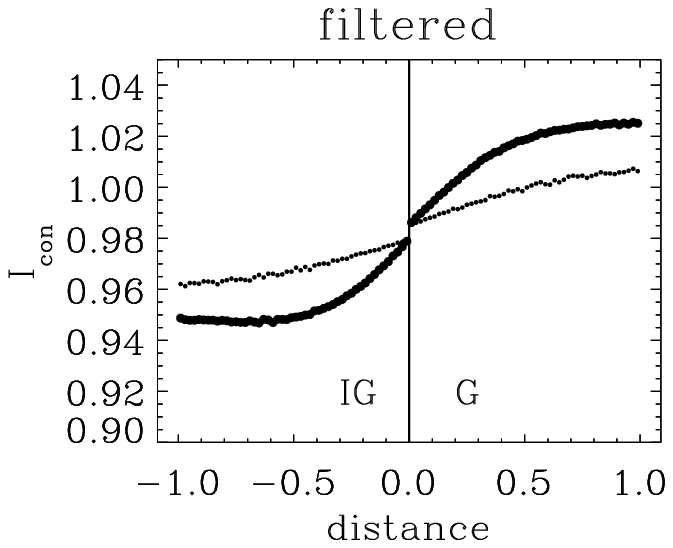}
\includegraphics[width=8cm,height=3.0cm]{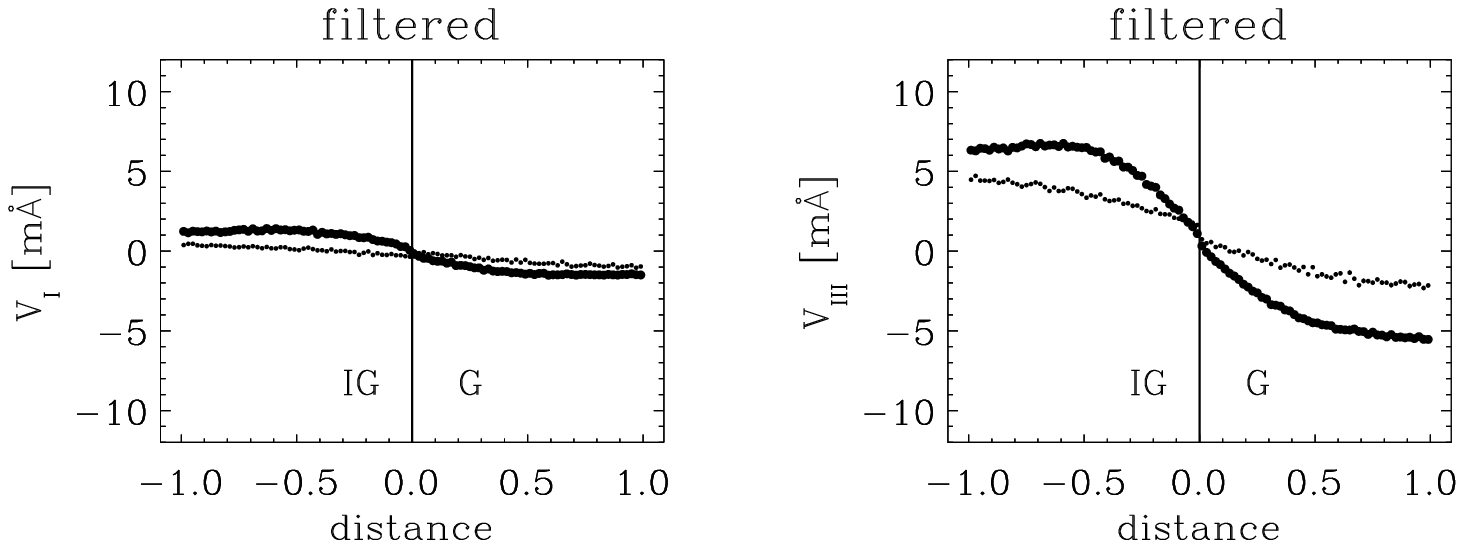}
\includegraphics[width=8cm,height=3.0cm]{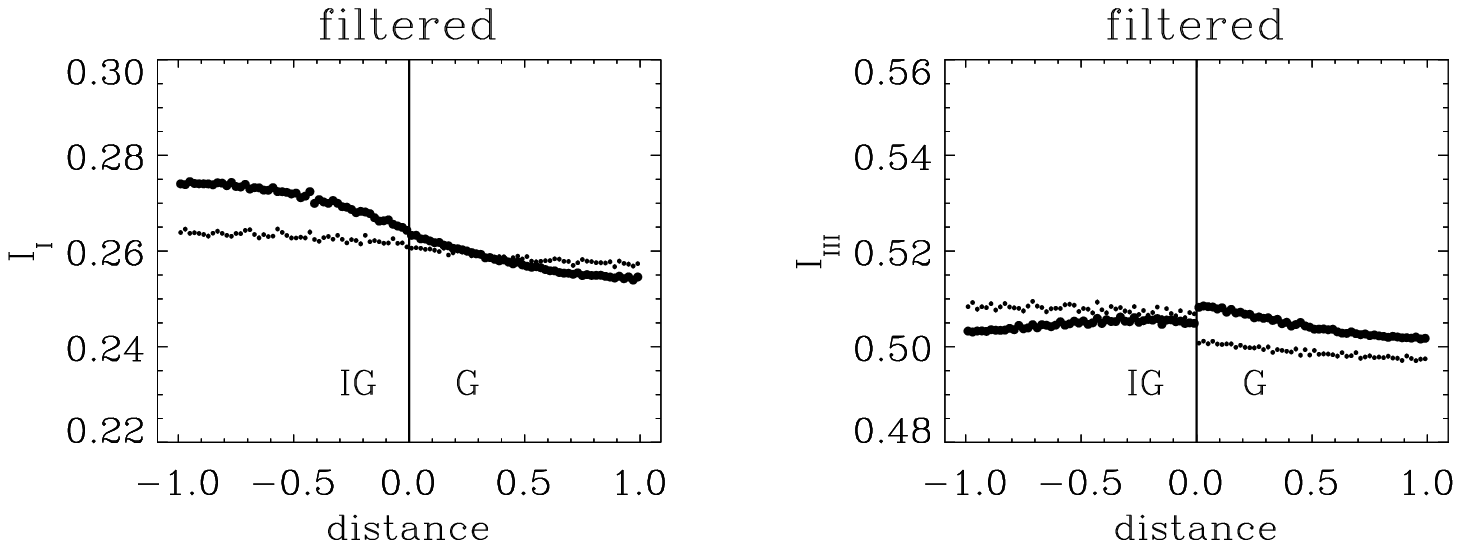}
\includegraphics[width=8cm,height=3.0cm]{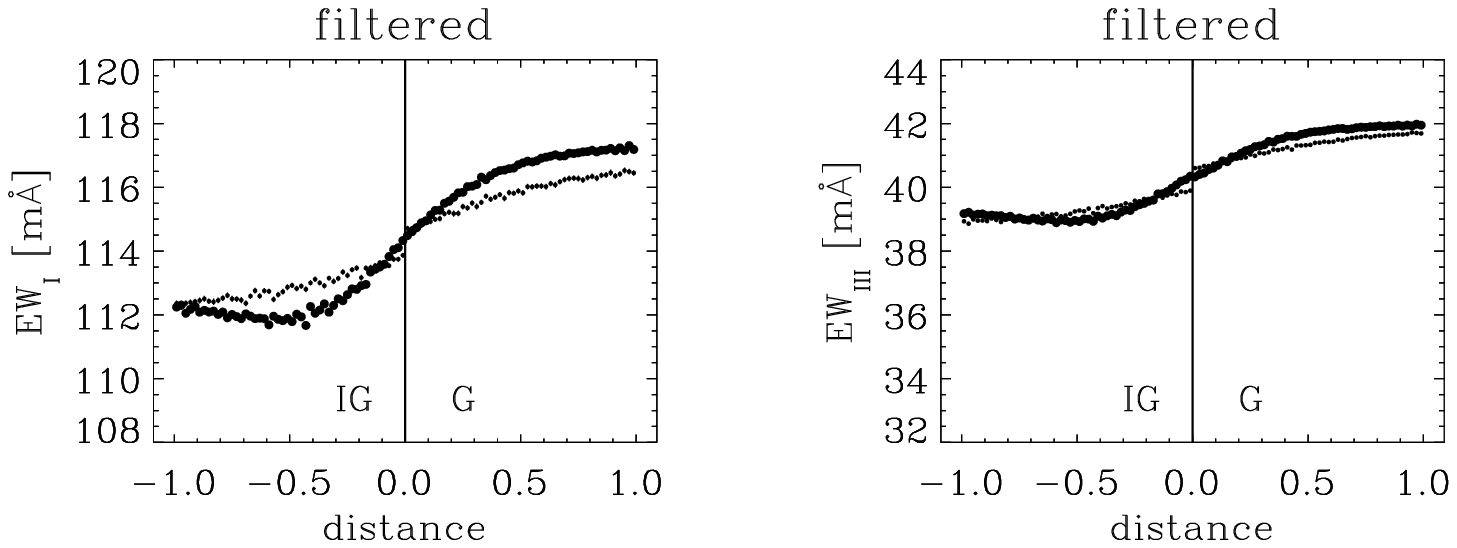}
\includegraphics[width=8cm,height=3.0cm]{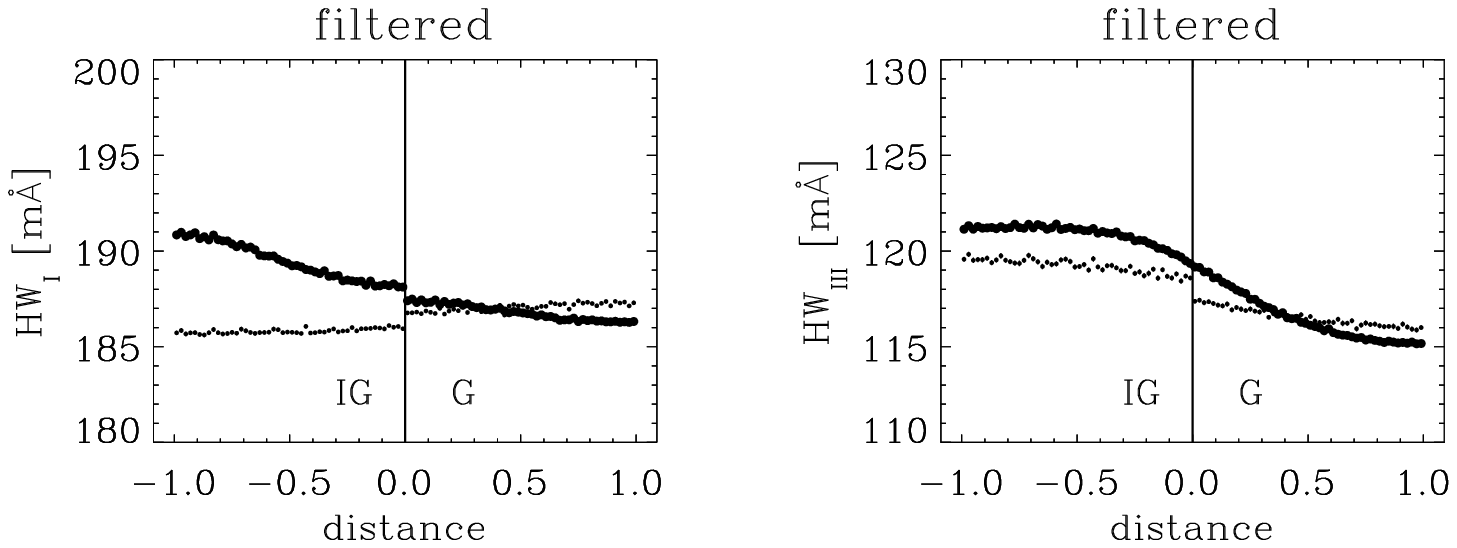}
\caption[]{$I_{\rm con}$, $V$, $I$, $EW$ and $HW$ {\it vs.} normalised distance to the granular border for small and large granules (G) and intergranular lanes (IG) for the filtered images of both lines. Small structures (0\,\farcs4 $<$ sizes $<$ 1\,\farcs4) are represented by thin points and large structures (1\,\farcs4 $<$ sizes $<$ 4\,\farcs0) by thick ones.}
\label{dist1}
\end{figure}

Figure \ref{dist1} shows the averaged values of the different line parameters for each bin of 0.02 normalised distance.

\noindent
We have split the data sets into small (sizes $<$ 1\,\farcs4) and large structures (1\,\farcs4 $<$ sizes $<$ 4\,\farcs0). The results correspond to images filtered of p-modes. All time steps are included in these computations and their results are averaged.\\

\noindent
From these computations one can obtain the following information: 

\begin {itemize}
\item[$\bullet$] $I_{\rm con}$:
For the family of large granular cells the curves show a sort of intensity level of  saturation in both the centre of the granules and of the intergranular lanes. The rest of the transition can be described as a monotonic increase toward
the granule. In case of small granular cells the general trend is essentially
linear. The discontinuity (jump) at the granular border (also visible for other line parameters) could be only apparent as a consequence of the uncertainty in the positions of the inflection points, which are not determined with sub-pixel precision. A glance at the upper panel of Fig. \ref{dist1} reveals that the
average of intensity values within each panel is not equal to 1. This is
caused by the fit of our spectra to the local continuum of the Li\`ege-Atlas (Delbouille et al. \cite{delbouille73}) which do not correspond to the global continuum of this spectral region.\\
\item[$\bullet$] $V$:
We find positive values (red shift) in the intergranular lanes and
negative values (blue shift) in the granules as expected. The vertical
speeds show increasing blue (red) shifts towards the centre of the granules
(intergranular lanes). This is not in contradiction with findings of other authors indicating that upflows in large granules peak adjacent to the intergranular lanes; this refers to particular cases of very large granules whereas in our average of large granules sizes ranging from 1\,\farcs4 to 4\,\farcs0 are included. Like in the case of $I_{\rm con}$ the curves of
$V_{\rm III}$ for large granular cells present a saturation level of the
vertical speed at the centre of the granules and the intergranular lanes and
the strongest gradient is close to the granular borders. We observe an
attenuation of the velocities with height in the photosphere that is in
agreement with the literature and the results obtained before from the computation of response functions and correlation coefficients.\\   
\item[$\bullet$] $I$:
For the higher formed line (Line I) we find larger values of $I$ in
the intergranular lanes, confirming the reversal of temperature in higher
layers. The line core intensities increase for both kinds
of granular cells almost linearly from the centre of the granule to the central
part of the intergranular lane. For Line III the difference of line core
intensities between granules and intergranular lanes for small and large
structures is very small, which confirms the previous results obtained by
correlation and coherence analysis. We cannot expect large temperature
fluctuations for line core intensities of this line, because most of the
convective information related to $I_{\rm III}$ comes from layers around
$\log \, \tau$ = $-$1.0, where the inversion of temperature occurs (see
Table \ref{maxima1} and Fig. \ref{rfresvel}a). Note, however, that in Line III 
the global trend for the case of small granular cells is steeper than that for
large granular ones, whereas in case of Line I the opposite occurs.\\
\item[$\bullet$] $EW$:
The equivalent width (see Fig. \ref{dist1}) is larger in the granules as
compared with the intergranular lanes. It shows a minimum located at the middle of the distance between the intergranular lane and the border, followed by a monotonic increase and a stabilisation towards the granular centre. Following Gadun et al. (\cite{gadun97}) the steeper temperature gradient in hot granular areas always produces deeper line profiles and respectively larger $EW$.\\
\item[$\bullet$] $HW$:
A decrease of $EW$ only caused by a change of temperature should also produce an decrease of $HW$. However, in Fig. \ref{dist1} the opposite behaviour can be observed (small granular cells in Line I are an exception), a decrease of $EW$ corresponds to an increase of $HW$. The simplest way to explain that is to include again the ingredients of turbulence or strong velocity gradients. Thus, the enhancement of $HW$ can be interpreted as caused by an enhancement of turbulence from the border of the granule to the intergranular lane. In case of Line I, the smoother increase of $HW$ throughout the intergranular lane could be justified as a partial compensation by temperature changes, that is also reflected in a strong decrease of $EW$. Nevertheless, the absolute peak to peak variation ($\sim$ 8 m\AA) remains similar for both lines. These results are not in complete agreement with Nesis et al. (\cite{nesis93}, \cite{nesis99}), who claim that in the deep photosphere, enhanced turbulence is concentrated predominantly near granular borders, while at higher layers the turbulence spreads out over the entire intergranular space. 
According to Solanki et al. (\cite{solanki}) the increase of $HW$ can be justified as a consequence of the vertical velocity structure of the atmosphere in intergranular lanes, and turbulence must not necessarily be invoked. 
\end{itemize}

\section{Detailed analysis of the variation of line parameters in space and time}

\begin{figure}[h]
\centering
\includegraphics[width=8cm,height=7cm]{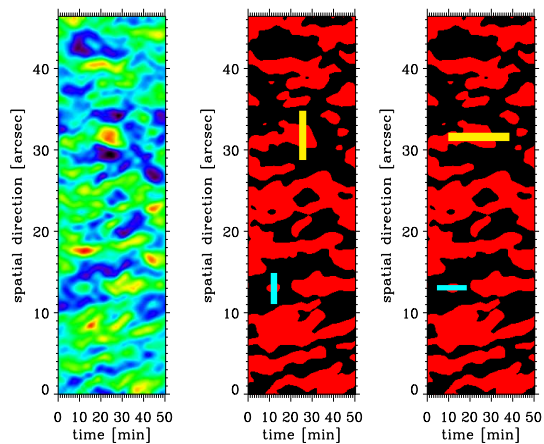}
\caption[]{False colour image (left panel) and binary images of $I_{\rm con}$ representing the selected areas for scans across small (small markers) and large granular cells (larger markers) in spatial (middle panel) and temporal (right panel) direction.}
\label{spasmselect}
\end{figure}

In this section detailed studies about the variation of
line parameters in space and time, separately for a small and a large
granular cell (see Fig. \ref{spasmselect}) by averaging in sampled time or space, are represented.

\begin{figure}
\centering
\includegraphics[width=8cm,height=10cm]{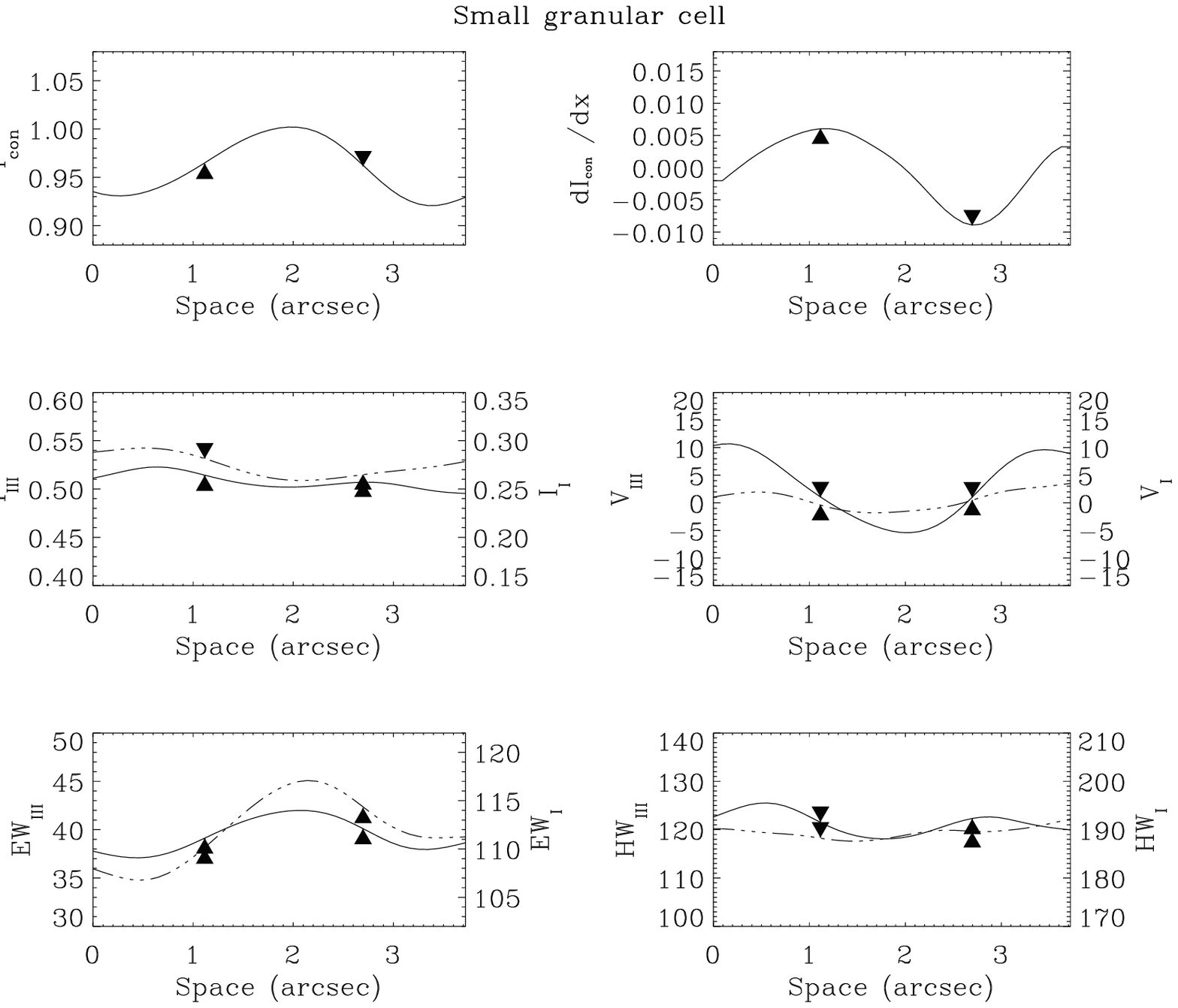}
\caption[]{Convective component: Spatial scan across a small granular cell, where in the four lower panels Line III is represented by solid lines, Line I by dashed ones. Granular borders are indicated by arrows.}
\label{spasm}
\end{figure}

\begin{figure}
\centering
\includegraphics[width=8cm,height=10cm]{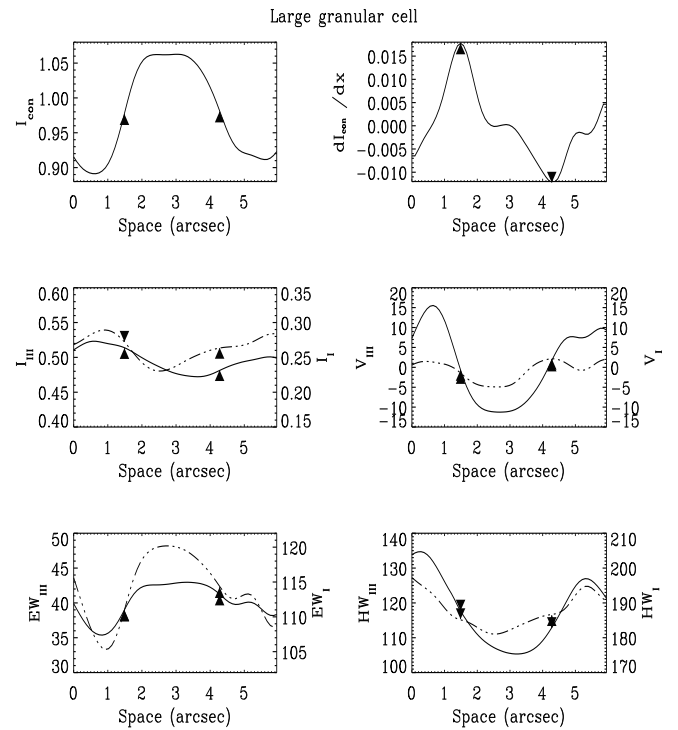}
\caption[]{Convective component: Spatial scan across a large granular cell, where in the four lower panels Line III is represented by solid lines, Line I by dashed ones. Granular borders are indicated by arrows.}
\label{spasm1}
\end{figure}

Figures \ref{spasm} and \ref{spasm1} show the spatial variation of line parameters across a small (size $<$ $1\,\farcs$5) and a large (size $\sim$ $3\,\farcs$0) granular cell, respectively. The borders of the granules are defined as the inflection points of $I_{\rm con}$, i.e. the extrema of the spatial gradient of intensities. We find the variation of line parameters more pronounced across the large granular cell. For $V$ and $HW$ this variation is attenuated with increasing height. Thus, these particular examples fit very well to the observed mean trend for $V$ and $HW$ in Fig. \ref{dist1}. $I_{\rm con}$ and $V$ show a broad maximum at the granular centre of the large cell. In lower layers the profile of the variation of vertical velocities in both cells shows a good (negative) correlation with the variation of $I_{\rm con}$, whereas, as expected, this correspondence weakens for higher layers (see Table \ref{ccc}). In the larger structure, the minima of the variation of the line core intensity are shifted towards the granular borders. In concordance with our results in Sect. \ref{dep}, higher (lower) values of equivalent width (full width at half maximum) above the granule (intergranular lane) are found.\\

The temporal evolution of the several line parameters has also been studied for both cells (see Fig. \ref{spasmselect}). The maximum of the $I_{\rm con}$ variation coincides in time with the extrema of the remaining line parameters. 
 
\section{Conclusions}
The vertical structure of the solar photosphere has been investigated, analysing the global and spatial variation of line parameters obtained from a time series of 1D spectra including two \ion{Fe}{i} lines with well separated formation heights. Thus we have been able to study how granules overshoot from the convection zone into the stable radiative atmosphere. For a qualitative interpretation of our results the formation heights of the several line parameters have been estimated, using the concept of the response function. The images, mapping the values of different line parameters in space and time have been filtered of p-modes. \\

The convective motions of structures larger than $1\,\farcs0$ penetrate up to the highest layers of the photosphere ($\sim$ 435 km). This conclusion stems from the significant correlation and coherence between continuum intensity and vertical velocities for both lines. In case of the higher forming line (Line I) the correlation and the coherence decrease which can be interpreted as a less efficient penetration of convective motions into the highest layers. In concordance, a significant correlation between the vertical velocities of both lines is found. This correlation is larger before filtering the p-modes, which implies a significant contribution of the oscillatory velocities at both layers. 

Regarding the intensity pattern, in very high layers ($\sim$ 435 km) only structures larger than $2\,\farcs0$ are connected with those at the continuum layer, but showing inverted brightness contrast. Thus the existence of an inversion of temperature is confirmed. We find such an inversion layer at $\sim$ 140 km, coinciding with the height of formation of the line core intensity of Line III. This is derived from the fact that neither significant correlation nor coherence of the line core intensity ($I_{\rm III}$) with continuum intensity ($I_{\rm con}$) and line core velocity ($V_{\rm III}$) is found. 

The phase shifts found between $V_{\rm I}$ and $I_{\rm con}$, $I_{\rm I}$, $V_{\rm III}$ could be interpreted as a possible evidence of gravity waves superposed on the convective motion (Deubner \cite{deubner89}) in higher layers.\\

From the study of the dependence of line parameters to the distance from the granular borders, for both lines an enhancement of full width at half maximum ($HW$) towards the intergranular lane is found. This goes together with a decrease of equivalent width ($EW$). A decrease of $EW$ only caused by a change of temperature should produce a decrease of $HW$. The simplest way to explain our findings is to include the ingredients of turbulence or strong velocity gradients in the intergranular lanes.

For the continuum brightness and vertical velocities for the family of large granular cells (1\,\farcs4 $<$ sizes $<$ 4\,\farcs0) a level of saturation in both, the centre of the granules and of the intergranular lanes is found. This reflects only an average behaviour and it does not necessarily correspond to cases of individual large granules, reported by several authors (e.g. Nesis et al. \cite{nesis92}, \cite{nesis93}, \cite{nesis97}; Hirzberger \cite{hirzberger02}) indicating that upflows in large granules peak adjacent to intergranular lanes.\\

The study of the variation of line parameters across two particular granular cells, small and large, respectively, confirms the mean behaviour, derived from the statistical analyses performed for the whole sample.\\

{\it Future aspects:}
In forthcoming papers, we will present studies of the spatial and temporal variation of physical parameters obtained from the inversion of our data using the inversion technique SIR of high resolution time series of spectroscopic data (Ruiz Cobo \& del Toro Iniesta \cite{ruizcobo92}, \cite{ruizcobo94}).  \\

\acknowledgements{We thank for the support by the Austrian Fond zur F\"orderung der wissenschaftlichen Forschung (Proj. Nr. 9184-PHY). This work was partially supported by the Spanish DGES under project 95-0028 and the Austrian-Spanish "Acciones Integradas" (grant no. 97-0028). Furthermore partial support by the Spanish Ministerio de Ciencia y Tecnolog\'\i a and by FEDER through project AYA2001-1649 is gratefully acknowledged. The Vacuum Tower Telescope is operated by the Kiepenheuer Institut f\"ur Sonnenphysik at the Spanish Observatorio
del Teide of the Instituto de Astrof\'\i sica de Canarias.}

\end{document}